\pdfoutput=1

\documentclass[11pt,a4paper]{article} 
\usepackage{jcappub} 

\usepackage{amsmath}
\usepackage{amsfonts}
\usepackage{graphicx}
\usepackage{bm}

\usepackage{varioref}
\usepackage{amssymb}
\usepackage{comment} 

\renewcommand{\topfraction}{0.99}

\title{Multifield consequences for D-brane inflation}
\author{Mafalda Dias,}
\author{Jonathan Frazer}
\author{and Andrew R. Liddle} 
\affiliation{Astronomy Centre, University of Sussex, Brighton BN1 9QH,
United Kingdom}

\emailAdd{m.dias@sussex.ac.uk}
\emailAdd{j.frazer@sussex.ac.uk}
\emailAdd{a.liddle@sussex.ac.uk}

\abstract{We analyse the multifield behaviour in D-brane inflation when contributions from the bulk are taken into account. For this purpose, we study a large number of realisations of the potential; we find the nature of the inflationary trajectory to be very consistent despite the complex construction.  Inflation is always canonical and occurs in the vicinity of an inflection point. Extending the transport method to non-slow-roll and to calculate the running, we obtain distributions for observables. The spectral index is typically blue and the running positive, putting the model under moderate pressure from WMAP7 constraints. The local $f_{\rm NL}$ and tensor-to-scalar ratio are typically unobservably small, though we find approximately $0.5\%$ of realisations to give observably large local $f_{\rm NL}$. Approximating the potential as sum-separable, we are able to give fully analytic explanations for the trends in observed behaviour. Finally we find the model suffers from the persistence of isocurvature perturbations, which can be expected to cause further evolution of adiabatic perturbations after inflation. We argue this is a typical problem for models of multifield inflation involving inflection points and renders models of this type technically unpredictive without a description of reheating.}


\begin{document}

\maketitle


\section{Motivation}

Inflation (for recent reviews see Ref.~\cite{infrev}) is widely viewed as the most elegant paradigm to understand the very early universe, but despite being a simple set-up, it is hard to fully describe it within an ultra-violet complete theory. The main reason for this is that its dynamics are highly sensitive to Planck-scale physics.
Generally speaking, Planck-suppressed contributions arise by integrating out heavy fields, which are present as extra degrees of freedom necessary for the ultra-violet completion of a theory. Unless protected by some specific symmetry, the inflaton will couple to these heavy fields, which results in a radiative instability of its mass and of the flatness of the potential. It is therefore of enormous interest to try to study inflation in an effective action that takes into account contributions of high energy physics. 


In string theory, the important degrees of freedom to take into consideration are the heavy moduli that arise from stabilized compactifications of the extra dimensions. To compute the detailed contributions that their coupling to the inflaton induces in the effective action requires full knowledge of the stabilized compactification, which is rarely possible. In this sense, it is important to identify and work with string set-ups that present a sufficient level of computability. This is the main motivation to look at inflation arising from the dynamics of D3-branes in warped throats. This scenario is not expected to be generic but it allows some major simplifications that make the task of building the effective action more achievable. 

The flux compactification causes warping of the manifold, giving rise to regions of the bulk with warped throats. Inflation can occur in this scenario when a D3-brane, corresponding to our four-dimensional space-time, is Coulomb attracted to an anti-D3-brane that sits at the infra-red tip of the throat, where it minimizes its energy. Inflation is driven by the dynamics of the D3-brane, and it behaves like a multi scalar field system, where the fields can be viewed as the physical coordinates separating the branes. The inflationary epoch ends when the branes collide and annihilate. 

What is special about these warped throat regions is that they can be approximated by a finite region of a non-compact conifold geometry, for which the metric and background fluxes are well known. This finite segment is then glued to the compact bulk at some ultra-violet scale. Corrections to this non-compact approximation will arise from the effect of fields on the throat, like stabilized moduli of the compact bulk, and are clearly examples of the Planck-suppressed contributions to the effective action mentioned above. 

Even working in this `simple' set-up, the effective action cannot be fully computed. At most, it is possible to calculate the form of the Planck-suppressed contributions, but they will come necessarily with an unknown Wilson coefficient. For this reason, to study the inflationary potential that arises from the D-brane scenario one needs to sample over a lot of realisations of random Wilson coefficients. Observational predictions need to be understood in the light of this statistical nature.\footnote{This issue is analogous to the challenge of making predictions in models of the string landscape. In order to compare predictions from this model with observation we must assume we are a typical observer. Of what class of observer we are typical is however a very difficult question to address and brings with it an inherent measure problem. We make no attempt to address this interesting challenge here, but it should be noted that a resolution of this problem will add a weighting to the distributions we present in this work.}

Another issue to keep in mind is that the conifold approximation for the throat does not hold at the infra-red tip. This implies that, if using such an approximation, the dynamics can only be analysed before the tip,
ignoring everything that occurs around and just before the collision of the branes. One can hope that this regime will not significantly affect the curvature perturbation $\zeta$, otherwise the predictions made are useless for comparisons with observations at decoupling time or after. 
To check if this is a reasonable assumption, it is necessary to keep track of isocurvature modes as they transfer power to the curvature perturbations. If they have not completely decayed by the end of the analysis, the curvature perturbation will continue to evolve as the brane moves into the tip. The inflationary trajectory is said to have not yet reached its `adiabatic limit'. The observable properties of $\zeta$ will then depend on unknown details of the tip, including reheating, making the model incomplete.

In this paper, our aim is to exhaustively study the possible D-brane dynamics above the tip and understand the consequent inflationary behaviour. For this purpose, we follow the most sophisticated set-up in the literature \cite{Agarwal}. This includes the sampling over Wilson coefficients and assumptions like the ones described above. The throat is approximated by a conifold parameterized by one radial and five angular directions that can be effectively viewed as the scalar fields driving inflation; this approximation only holds for a fixed range in the radial direction and it can be shown that within it DBI effects can be neglected. The framework and details of the construction of the potential are reviewed in~\S\ref{sec:D-brane}.

Since the potential that describes the motion of the D-brane is sensitive to all 6 coordinates, multifield effects have a profound impact on the dynamics of the inflationary trajectory and consequent curvature perturbations $\zeta$. To compute observables within this multifield superhorizon dynamics, we use the \emph{transport equations} method originally introduced in Refs.~\cite{MT1,MT2} along with an extension to non-slow-roll that will be described fully in a forthcoming paper \cite{us}. This technique is a realisation of the separate universe assumption in which the values of correlation functions of $\zeta$ and their tilts can be directly transported from horizon-crossing to the desired time of evaluation, as described in~\S\ref{sec:transport}. Since this is framed in terms of ordinary differential equations, it allows for a clean and efficient numerical implementation as well as a means of explicitly keeping track of the evolution of isocurvature modes.

As a consequence of our analysis, we obtain the probability of getting inflation, found to be in agreement with Ref.~\cite{Agarwal}, and the statistical distributions of observables predicted by our set-up. But the most interesting results of this work come from our understanding of the peculiarities of the multifield dynamics of the D-brane potential.  By looking in detail at the resulting inflationary trajectories, we are able to map trends in the distributions of observables to generic features of the potential. Moreover, we can see if these features allow the trajectories to reach their adiabatic limit. The outcome of our work is then more than the computation of predictions, being an investigation on the limitations of our set-up as a predictive and useful toy model for the D-brane scenario.

\section{D-brane inflation}
\label{sec:D-brane}
\subsection{D-branes in a warped throat}

As mentioned earlier, in the D-brane inflation scenario, our Universe lives in a D3-brane and experiences inflation due to its dynamics in a warped throat region of a stabilized compact space; the D3-brane is attracted by an anti-D3-brane that sits at the tip of the throat where its energy is minimized. When the two branes collide, they annihilate, inflation ends and reheating occurs \cite{Kachru, Cliff}.

\begin{figure}[t]
\centering
\includegraphics[width=10cm]{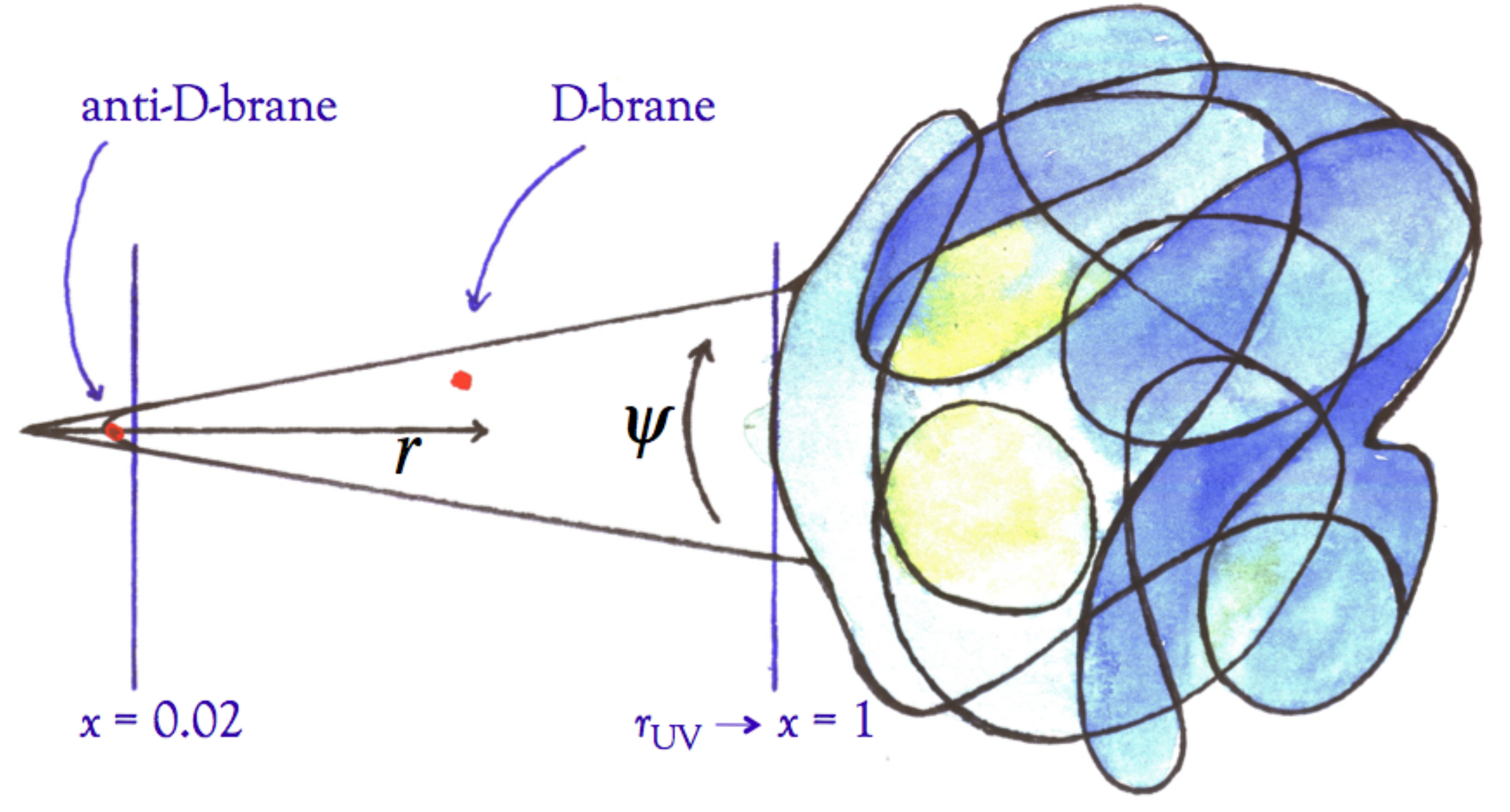}
\caption{The non-compact conifold approximation for the warped throat. This approximation holds between $x=1$, where the throat is glued to the compact bulk, and $x=0.02$, where the tip that cannot be described by our approximated geometry starts.}
\label{conifold}
\end{figure}

The throat region, excluding the tip, can be well approximated by a non-compact conifold geometry; the conical singularity that would arise at the tip is smoothed by fluxes such that the radial coordinate at this point is finite. For the purpose of this work, only the region above the tip will be considered. In this case, and ignoring logarithmic corrections to the warp factor, the background geometry can be approximated by:
\begin{equation}
ds^{2}=\left(\frac{R}{r}\right)^{2}g_{ij}d\phi^{i}d\phi^{j}= \left(\frac{R}{r}\right)^{2}\left(dr^{2}+r^{2}ds^{2}_{T^{1,1}}\right),
\end{equation}
where $r$ is the radial conical coordinate and $T^{1,1}$ is the coset space $\left(SU(2) \times SU(2)\right)/U(1)$ that describes the angular directions of the cone. The radius $R$ is approximately $r_{\rm UV}$, the coordinate at which the throat is glued to the compact bulk, as illustrated in Fig.~\ref{conifold}. In agreement with Ref.~\cite{Agarwal} we use the value $r_{\rm UV}=1$ throughout this work. It is useful to define a rescaled radial coordinate as $x \equiv r/r_{\rm UV}$ that in the cone region is always $0 \ll x < 1$. To ensure that the non-compact approximation always holds, we restrict our analysis to the regime comfortably above the tip, where  $0.02 < x < 1$. The value $x=0.02$ was chosen in agreement with Ref.~\cite{Agarwal}.

The $T^{1,1}$ space is parameterized by 5 angles $\Psi =\left\{ \theta_{1}, \theta_{2}, \varphi_{1}, \varphi_{2}, \psi \right\} $ where $0 \leq \theta_{1} \leq \pi$,  $0 \leq \theta_{2} \leq \pi$,  $0 \leq \varphi_{1} < 2\pi$,  $0 \leq \varphi_{2} < 2\pi$ and $0 \leq \psi < 4\pi$. \\

Throughout this paper we will use units $M^{-2}_{\rm Pl}=8 \pi G=1$. Also, in agreement with Ref.~\cite{Agarwal}, we use the value $r_{\rm UV}=1$. 
In this throat the D3-brane experiences a DBI inflationary Lagrangian like
\begin{equation}
{\cal L}=a^{3}\left(-T(\phi)\sqrt{1-\frac{T_{3}g_{ij}\dot{\phi}^{i}\dot{\phi}^{j}}{T(\phi)}}-V(\phi)+T(\phi) \right),
\end{equation}
where $a$ is the scale factor, $T_{3}$ is a constant representing the brane tension and, within the approximation where the logarithmic corrections to the warp factor can be ignored, $T(\phi)=T_{3}x^{4}$. The value of the warp factor at the tip is determined by the parameter $a_{0}$ such that $T(\phi)|_{\rm tip} \equiv T_{3}a_{0}^{4}$. Some physical arguments concerning the consistency of the set-up enforce a limit on how small $T(\phi)$ can get \cite{Baumann}; in this work, following Ref.~\cite{Agarwal}, we use the values $T_{3}=10^{-2}$ and $a_{0}=10^{-3}$. 

As mentioned in Ref.~\cite{Agarwal}, for our specific realisations of the D-brane action, the brane velocity is always very small compared to $T(\phi)$, making 
\begin{equation}
\label{DBI}
\frac{T_{3}g_{ij}\dot{\phi}^{i}\dot{\phi}^{j}}{T(\phi)} \ll 1.
\end{equation}
This is equivalent to saying that DBI effects are negligible, as we can rewrite the Lagrangian as
\begin{equation}
{\cal L}=a^{3}\left(\frac{1}{2}T_{3}g_{ij}\dot{\phi}^{i}\dot{\phi}^{j} -V(\phi) \right)
\end{equation}
and identify the canonical kinetic term rescaled by the constant $T_{3}$.

The fact that this simplification can be made is related not only to the choice of $T_{3}$ and $a_{0}$ but also to the fact that our analysis only includes the throat region above the tip. Condition~\eqref{DBI} breaks if $T(\phi)$ gets very small and, in fact, $T(\phi)$ decreases with $x$. Possible DBI inflation in this regime can have strong repercussions in the value of observables at the end of inflation; the inclusion of the whole throat in the computation of perturbations is then potentially very interesting but beyond the scope of this work.

\subsection{D-brane potential}
\label{sec:D-brane potential}

In the simplest form of this scenario, the potential that induces inflation has two contributions~\cite{Cliff}. First, there is the Coulomb interaction between the pair of branes, which is a multipole expansion where high-multipole terms are suppressed by powers of $a_{0}$. The leading terms are:
\begin{equation}
V_{C}=D_{0}\left(1-\frac{27D_{0}}{64\pi^{2}T_{3}^{2}r^{4}_{\rm UV}}\frac{1}{x^{4}}\right)
\end{equation}
where the parameter $D_{0} \equiv 2T_{3}a^{4}$ determines the overall scale of inflation.

Second, the coupling to the curvature induces, at leading order, a mass term like:
\begin{equation}
V_{M}=\frac{1}{3}\mu^{4}x^{2}
\end{equation}
where the scale $\mu^{4} \equiv D_{0}T_{3}r_{\rm UV}^{2}/M_{\rm Pl}^{2}$.

In this basic picture, inflation would actually be single-field, with the inflaton being the radial separation between brane and anti-brane, $x$. However, if the dynamics was determined by these two terms only, sufficient inflation couldn't be achieved \cite{Kachru}.
Such a potential has a single inflection point; this feature will be shown in the next sections to have strong consequences for the phenomenology of the full model.

This simplified picture ignores some important contributions to the potential experienced by the D-brane. One needs to take into account that the throat is finite and glued to a compact manifold, and as such, moduli stabilization from the bulk will necessarily have an impact on the throat geometry. These contributions can be viewed as corrections to the non-compact approximation, and will be denoted by $V_{\rm bulk}$.

Ideally, one would like to have the full knowledge of the 4-dimensional potential induced by the compactification flux on the brane dynamics, but this is not possible to achieve for a general Calabi--Yau bulk. However, it is known \cite{Baumann} that such a potential, in the non-compact background of the conifold, respects the Laplace equation:
\begin{equation}
\label{Laplace}
\nabla^{2} V_{\rm bulk}=0.
\end{equation}
Since we know completely the geometry of the conifold, this equation can be explicitly solved. We refer to these contributions, following the notation of Ref.~\cite{Baumann}, as the homogeneous contributions to $V_{\rm bulk}$.
 
Deviations from this expression, which holds for the non-compact background, can be obtained by allowing a source from the bulk. In this case, the Poisson equation looks like  \cite{Baumann}
\begin{equation}
\label{inhom}
\nabla^{2} V_{\rm bulk}=\frac{g_{s}}{96}\left|\Lambda\right|^{2},
\end{equation}
where $g_{s}$ is the string coupling constant and $\Lambda$ is proportional to the imaginary anti-self-dual three-form flux from the bulk. We will refer to these contributions as the inhomogeneous contributions to $V_{\rm bulk}$.
To solve this equation, a simplification can be used. Since these contributions are perturbations to the non-compact approximation, they can be assumed, up to a good approximation, to have the same structure as the homogeneous contributions. So the idea is to express them as an expansion of harmonic terms from the homogeneous solution. In other words, the solutions to the Laplace equation dictate the structure of the bulk contribution to the potential.\\

\textsc{Homogeneous contributions}:
The Laplace equation \eqref{Laplace} for our non-compact conifold can be written in the form of the expansion \cite{Klebanov}:
\begin{equation}
\label{Vhom}
V_{\rm hom \ bulk}(x,\Psi)=\mu^{4} \sum_{LM} C_{LM}x^{\Delta(L)}Y_{LM}(\Psi)
\end{equation}
where $C_{LM}$ are constant coefficients,  $Y_{LM}(\Psi)$ are the angular eigenfunctions of the Laplacian of the $T^{1,1}$ space and the subscripts $L\equiv \{l_{1},l_{2},R\}$ and $M \equiv \{m_{1},m_{2}\}$ represent the quantum numbers under the $T^{1,1}$ isometries. The powers $\Delta(L)$ are related to the eigenvalues of the Laplacian and are given by 
\begin{equation}
\Delta(L)\equiv -2 + \sqrt{6l_{1}(l_{1}+1)+6l_{2}(l_{2}+1)-3R/4+4}.
\end{equation}

The magnitudes of $C_{LM}$ are highly dependent on details of specific compactifications, so they need to be considered unknown parameters. Using the scale $\mu^{4}$, considerations of Ref.~\cite{Baumann2} suggest that $C_{LM} \sim {\cal O}(1)$, so a way to deal with this lack of knowledge is to scan randomly over values in this range.
To take the leading contributions of this term, one needs to consider the lower values of $\Delta(L)$. The maximum value desired for $\Delta(L)$ determines the truncation of the summation.\\

 \textsc{Inhomogeneous contributions}:
To solve Eq.~\eqref{inhom} as an expansion of the type of Eq.~\eqref{Vhom}, one needs to identify the radial scaling of the flux $\Lambda$ in terms of the quantum numbers $L$ and $M$ of $T^{1,1}$. It is possible to classify the flux in 3 different series, I, II and III, regarding their different radial scaling \cite{Baumann}. 
The radial scaling of $|\Lambda|^{2}$ is given by \footnote{Some contractions of flux series vanish following the considerations of Ref.~\cite{Baumann}.}
\begin{equation}
\Delta(L_\alpha, L_\beta)_{\rm inhom \ bulk} \equiv \Delta_\alpha(L_{\alpha})+\Delta_{\beta}(L_{\beta})-4
\end{equation}
where $\alpha$ and $\beta$ run over the 3 different series I, II and III and
\begin{equation}
\Delta_{I}(L)\equiv -1 + \sqrt{6l_{1}(l_{1}+1)+6l_{2}(l_{2}+1)-3R/4+4},
\end{equation}

\begin{equation}
\Delta_{II}(L)\equiv \sqrt{6l_{1}(l_{1}+1)+6l_{2}(l_{2}+1)-3R/4+4},
\end{equation}

\begin{equation}
\Delta_{III}(L)\equiv 1 + \sqrt{6l_{1}(l_{1}+1)+6l_{2}(l_{2}+1)-3R/4+4}.
\end{equation}
It is then possible to write the inhomogeneous contributions as:
\begin{equation}
V_{\rm inhom \ bulk}(x,\Psi)=\mu^{4} \sum_{L_{\alpha}M_{\alpha},L_{\beta}M_{\beta}} C_{L_{\alpha}M_{\alpha}L_{\beta}M_{\beta}} x^{\Delta(L_\alpha, L_\beta)}Y_{L_{\alpha}M_{\alpha}}(\Psi)Y_{L_{\beta}M_{\beta}}(\Psi).
\end{equation}
To write this expression with the same structure as Eq.~\eqref{Vhom}, the angular part needs to be expanded in terms of $Y_{LM}$ of $T^{1,1}$ as 
\begin{equation}
Y_{L_{\alpha}M_{\alpha}}(\Psi)Y_{L_{\beta}M_{\beta}}(\Psi) = \sum_{LM} A_{\alpha\beta}Y_{LM}(\Psi)
\end{equation}
such that finally,
\begin{equation}
V_{\rm inhom \ bulk}(x,\Psi)=\mu^{4} \sum_{L_{\alpha},L_{\beta}}\sum_{LM} C_{LM} x^{\Delta(L_\alpha, L_\beta)}A_{\alpha\beta}Y_{LM}(\Psi).
\end{equation}

The constants $C_{LM}$ correspond to the random parameters associated to each $Y_{LM}$ from the homogeneous contribution, and, just as in that case, the maximum value desired for $\Delta(L_{\alpha},L_{\beta})$ determines the truncation of the summation.\\

\begin{figure}[t]
\centering
\includegraphics[width=10cm]{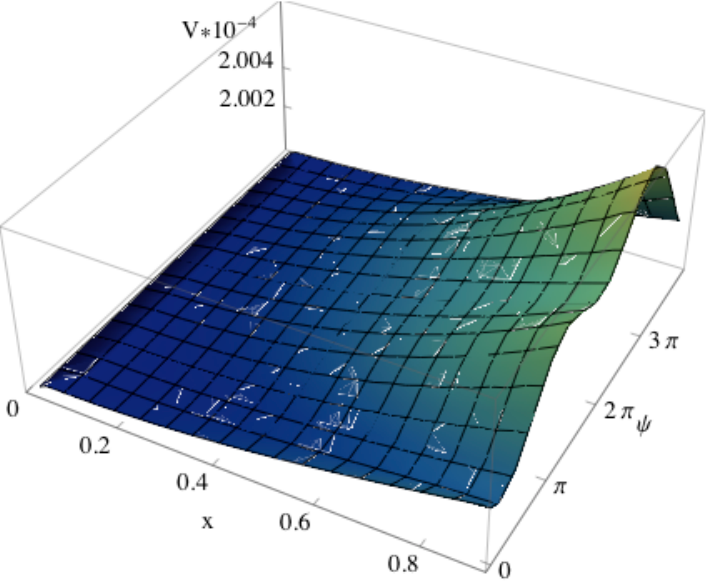}
\caption{A typical realisation of the D-brane potential with only one angular direction, $\psi$, active.}
\label{potential}
\end{figure}

The total potential experienced by the D-brane in the throat is then
\begin{equation}
\begin{split}
V(x,\Psi)&=V_{C}+V_{M}+V_{\rm hom \ bulk}+V_{\rm inhom \ bulk} 
\\&=D_{0}\left(1-\frac{27D_{0}}{64\pi^{2}T_{3}^{2}r^{4}_{\rm UV}}\frac{1}{x^{4}}\right) + \frac{1}{3}\mu^{4}x^{2} 
\\& +\mu^{4} \sum_{LM} C_{LM}x^{\Delta(L)}Y_{LM}(\Psi) + \mu^{4} \sum_{L_{\alpha},L_{\beta}}\sum_{LM} C_{LM} x^{\Delta(L_\alpha, L_\beta)}A_{\alpha\beta}Y_{LM}(\Psi)
\end{split}
\end{equation}
A specific realisation of this potential, with only one angular direction being taken into account, is shown in Fig.~\ref{potential}.

\section{Experimental procedure}
\label{sec:experiment}

Having established that the brane potential necessarily has a level of randomness to be able to encompass complex contributions, it is important to construct a useful sample of realisations for the study of the emergent inflationary behaviour. 

The first thing to specify is the maximum values of $\Delta(L)$ and $\Delta(L_\alpha, L_\beta)$ in the potential. Since the mass term has power $\propto x^{2}$, it makes sense to include at least all terms with $\Delta \leq 2$. For computational reasons, and in accordance with Ref.~\cite{Agarwal}, we looked at potentials with $\Delta \leq 3$ and with $\Delta \leq \sqrt{28}-3/2$. This corresponds to a total of 121 and 390 independent terms in the potential, respectively.\footnote{Note that these numbers differ slightly from those of Ref.~\cite{Agarwal}. The origin of this discrepancy lies in details of the expansions performed on $V_{\rm bulk}$.} The values taken by $\Delta$ are: $1,3/2,2,5/2,\sqrt{28}-5/2,3,\sqrt{28}-2,7/2,\sqrt{28}-3/2$. 

The second thing to decide is how to generate the random $C_{LM}$ coefficients. Following Ref.~\cite{Agarwal}, we define $C_{LM} \equiv   Q\hat{C}_{LM}$ such that $\hat{C}_{LM}$ is a distribution with unit variance, encapsulating the information on the distribution, and $Q$ is the root mean square size of $C_{LM}$, encapsulating the information on its magnitude. 

Drawing conclusion on the predictions of the model would be problematic if the inflationary behaviour emerging from our sample was dependent on the type of distribution of  $\hat{C}_{LM}$. 
Fortunately, as shown in Ref.~\cite{Agarwal}, this is not the case. In this work, we use a Gaussian distributed $\hat{C}_{LM}$. 

\begin{figure}[t]
\centering
\includegraphics[width=10cm]{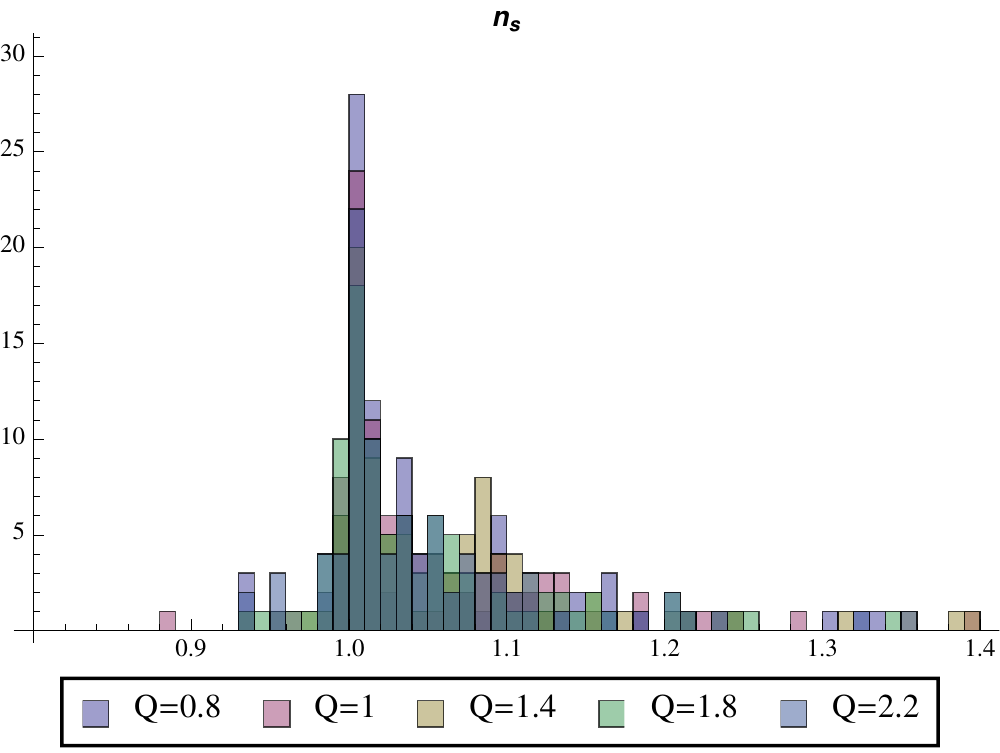}
\caption{Distributions for the scalar spectral index for the model with 2 active fields. The distributions were taken with different values of $Q$ and show very similar behaviours ($\Delta_{\rm MAX}=3$ in all distributions). }
\label{qtest}
\end{figure}

Regarding the choice of $Q$, a similar argument could be invoked. As mentioned in the previous section $Q \sim {\cal O}(1)$, but the probability of inflation is very sensitive to its precise value \cite{Agarwal}. If the emergent behaviour, given that inflation occurs, was also sensitive to this choice of $Q$, it would be hard to make general predictions.
To ensure that no issue would arise from this effect, we tested the dependence of the inflationary phenomenology on the choice of $Q$ and found it to be independent. This is demonstrated in Fig.~\ref{qtest}, where, as an illustrative example, the distribution of the spectral index for a two-field sample is plotted for five different values of $Q$.

The last thing to fix before constructing the sample is the choice of initial conditions, for which we follow Ref.~\cite{Agarwal} precisely. Since the potential is statistically invariant under angular translation in $T^{1,1}$, generating multiple realisations starting always at the same angular coordinate automatically encompasses the statistical effect of varying initial conditions within a single realization. We hence consider just one initial condition per realization, arbitrarily taken as the angular coordinates being $\Psi_{0}=\{1,1,1,1,1\}$. Regarding the initial radial direction, and following arguments from Ref.~\cite{Agarwal}, we chose $x_{0}=0.9$. We set all the initial velocities to zero, $\dot{x}_{0}=\dot{\Psi}_{0}=0$, leaving the study of the possible impact of initial velocities on inflationary phenomenology for future work.

We can now present our experimental procedure for the building of a statistical sample of inflationary trajectories. This procedure was first used in Ref.~\cite{TegmarkInf} and more recently in Refs.~\cite{me1,me2,Agarwal}:
\begin{enumerate}
\item Generate a random potential $V(x,\Psi)$ starting at $(x_{0},\Psi_{0})=(0.9,1,1,1,1,1)$ and evolve to find the field trajectory.
\item If the model gets stuck in eternal inflation, \textit{i.e.} does not reach $x=0.02$, reject.
\item If the brane gets ejected from the throat, \textit{i.e.} $x$ gets larger than 1, reject. 
\item Once the brane has reached $x=0.02$, if the number of e-folds of inflation $N<60$, reject, as insufficient inflation occurred, otherwise calculate observables. 
\item Repeat steps 1-4 many times to obtain a statistical sample.
\end{enumerate}
Several sets of samples were generated by changing some parameter, either $\Delta_{\rm MAX}$, $Q$ or the number of active fields. When mentioning two-field samples, we are referring to a model with the radial direction and one angular direction active. When we discuss more active fields, we are referring to models with additional angular directions. 

The probability of achieving successful inflation in this set-up is in agreement with Ref.~\cite{Agarwal}. If $Q$ is too small, most of the trajectories do not produce 60 e-folds of inflation; if $Q$ is too large, most of the trajectories lead to an ejection of the brane. The optimum value of $Q$ lies between these two regimes.
As the number of active fields increases, just as when $\Delta_{\rm MAX}$ increases, $P(N>60)$ becomes more sensitive to the choice of $Q$, and decreases slightly.  Illustrative values are shown in Table~\ref{t:table}. It is interesting to note that even for the optimum $Q$, more than half of the rejections are due to the brane being ejected from the throat, in agreement with Ref.~\cite{Agarwal}. 

\begin{table}[h]
\centering
\begin{tabular}{|c|c|c|c|c|c|c|c|c|c|c|c|c|}
    \hline
                      
 Fields         & 2  & 2  & 3  & 3 & 4 & 6    \\ \hline
$ \Delta_{\rm MAX} $& 3  & $\sqrt{28}-3/2$ &  3 & $\sqrt{28}-3/2$   &  3 & 3  \\ \hline
$P(N>60)$     & $3 \times 10^{-4}$   & $3 \times 10^{-4}$ & $2 \times 10^{-4}$ & $9 \times 10^{-5}$  & $6 \times 10^{-5}$ & $3 \times 10^{-5}$   \\ \hline 
    \end{tabular}
\caption{Probability of getting successful trajectories ($N>60$) as a function of the number of active fields and $\Delta_{\rm MAX}$.}
\label{t:table}
\end{table}

\begin{figure}[t]
\centering
\includegraphics[width=15cm]{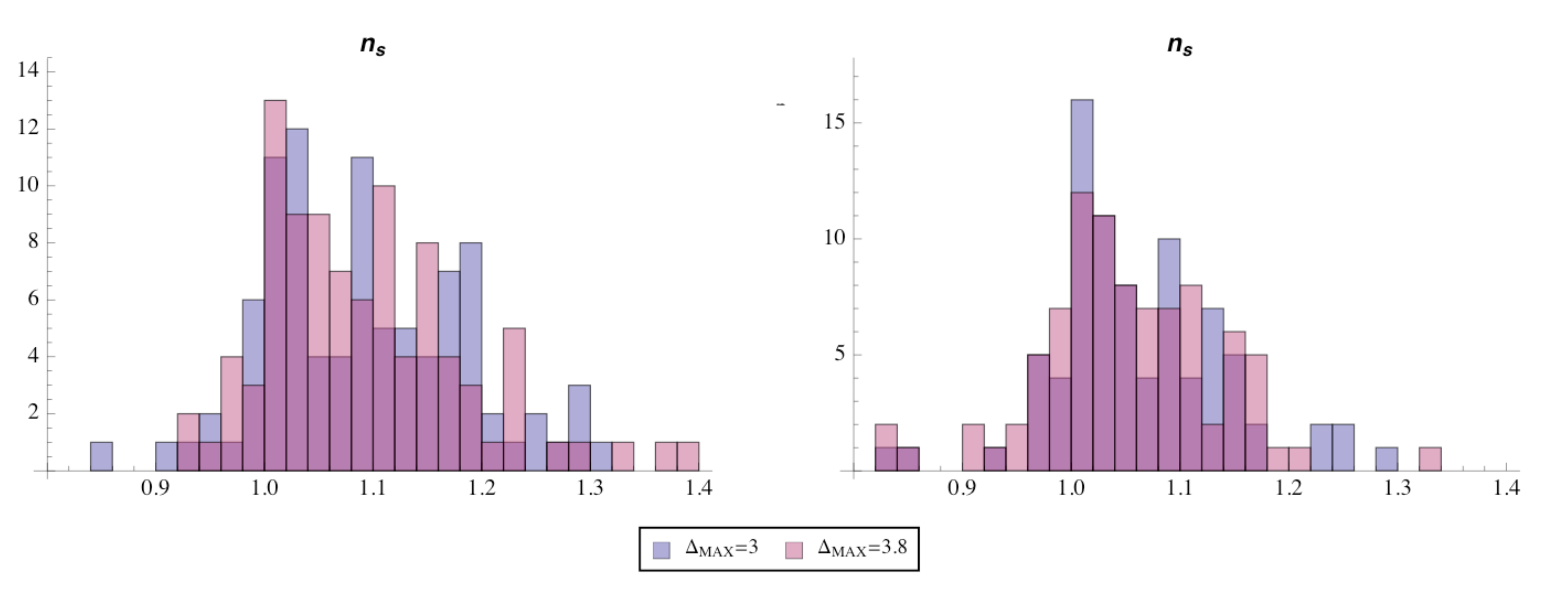}
\caption{Distributions for the scalar spectral index for the model with 2 active fields, left, and 3 active fields, right. The distributions were taken with $\Delta_{\rm MAX}=3$ and $\Delta_{\rm MAX}=\sqrt{28}-3/2 \approx 3.8$ and show very similar behaviours ($Q$=1.4 in all distributions).}
\label{deltatest}
\end{figure}

Regarding the value of $\Delta_{\rm MAX}$, we noticed that the inflationary phenomenology does not depend strongly on the chosen value. An illustrative example is shown in Fig.~\ref{deltatest}, for the spectral index of models with two and three fields. For this reason, in what follows, we concentrate on distributions with $\Delta_{\rm MAX}=3$.

In the next section, we present the techniques used for the computation of observables.

\section{Computing the curvature perturbation}
\label{sec:transport}

A central feature of any model of inflation with more than one active scalar field is that the primordial curvature perturbation $\zeta$ evolves on superhorizon scales. To compute this, almost all methods in the literature to date make use of some variant of the separate universe assumption~\cite{RS1,RS2,YST}. The idea is to understand the final curvature perturbation as the result of the scatter of a collection of equal size space-time patches, within which the field values at a given time are constant. Let us make this more precise.

\subsection{The separate universe assumption is a geometrical optics approximation}

The separate universe assumption states that, when smoothed on some physical scale $L$ much larger than the horizon scale, the average evolution of each $L$-sized patch can be computed using the background equations of motion and initial conditions taken from smoothed quantities local to the patch. The evolution of $\zeta$ can be understood as the  variation in the expansion of these patches.
Since each smoothed patch corresponds to a position in phase space, the evolution of $\zeta$ can be determined by the evolution of an ensemble of points in the classical phase space. These points are subject to the laws of statistical physics and hence evolve according to the Liouville equation. 

Under the separate universe assumption, interactions between patches are negligible and therefore all that is required  is a mapping of the initial conditions to a final state. The final distribution in phase space can then be viewed as an \emph{image} of the initial conditions. This mapping simply follows a flow generated by the background theory and can be calculated in precisely the same way as geometrical optics enables us to calculate the image generated by a source of light rays. This optical description was made precise in Ref.~\cite{optics}, which we briefly summarise in this section, referring the reader to Ref.~\cite{optics} for a more detailed discussion.

%
%
%
\subsection{Transport equations}

Since slow-roll approximations are not valid in general, we are required to work in a $2N_{\rm F}$-phase space. This consists of $N_{\rm F}$ fields $\phi_{i}$ as well as their momenta $p_{i}\equiv \phi'_{i}$, where primes represent differentiation with respect to the number of e-folds $N$. The fields $\phi_{i}$ and $p_{i}$ are treated on an equal footing, so from now on we will denote a point in phase space by $\varphi_{\alpha} \equiv \left\{ \phi_{i}, p_{i} \right\}$ where $\alpha$ runs from 1 to $2N_{\rm F}$.

In canonical models of inflation, if we set initial conditions near horizon-crossing, the initial distribution of field perturbations will be close to Gaussian \cite{22}. Furthermore, typical spacing between arbitrarily selected members of the ensemble is of order the quantum scatter. It follows that the trajectories traversed by the ensemble trace out a narrowly-collimated spray or `bundle' of rays in phase space with an initial Gaussian distribution. This scenario is well studied in the optics literature since many lasers have this characteristic.

Cross-sections within the bundle of trajectories may be focused, sheared or rotated by refraction. It is ultimately through these distortions that any evolution in $\zeta$ occurs. To describe these distortions quantitatively, 
it is only necessary to know how some basis which spans the cross-section is transported from slice to slice. Denoting the difference between two field values at equal-time positions $\bf{x}$ and $\bf{x}+\bf{r}$ by $\delta\varphi_{\alpha}(\bf{r})$, we have an appropriate basis. This basis evolves along the beam as \cite{optics}
\begin{equation}
		\frac{d \delta \varphi_\alpha({\bf{r}})}{d N}
		=  u_{\alpha\beta}[\varphi({\bf{x}})] \delta \varphi_\beta({\bf{r}})
		+ \frac{1}{2} u_{\alpha\beta\gamma}[\varphi(\bf{x})]
			\delta \varphi_\beta(\bf{r})
			\delta \varphi_\gamma({\bf{r}})
		+ \cdots .
	\label{eq:real-deviation}
\end{equation}
where $u_{\alpha \beta} \equiv \partial_\beta u_\alpha$ is the expansion tensor defined as the derivative with respect to the fields of the background flow $u_{\alpha}[\varphi(\bf{x})] \equiv \varphi_{\alpha}'(\bf{x})$ and similarly  $u_{\alpha \beta \gamma} \equiv \partial_\gamma u_{\alpha \beta}$. For clarity, we will drop the explicit $\varphi(\bf{x})$ dependence from now on.

The expansion tensor can be decomposed as a dilation $\theta =  {\rm tr}\, u_{\alpha \beta}$, a traceless symmetric shear $\sigma_{\alpha \beta}$, and an antisymmetric twist $\omega_{\alpha \beta}$,
\begin{equation}
		u_{\alpha \beta} \equiv \frac{\theta}{2N_F} \delta_{\alpha \beta}
			+ \sigma_{\alpha \beta} + \omega_{\alpha\beta} ,
		\label{eq:expansion-tensor}
	\end{equation}
Dilation describes a rigid, isotropic rescaling of $\delta \varphi_\alpha$ by $1+\theta$, representing a global tendency of the light rays to focus or defocus. The shear $\sigma_{ij}$ represents a tendency for some light rays within the beam to propagate faster than others. The twist $\omega_{\alpha\beta}$ describes a tendency of neighbouring trajectories to braid around each other. 

The observables of interest, like the power spectrum of $\zeta$, its spectral index $n_{s}$, the local non-Gaussianity parameter $f_{\rm NL}$, etc, are related to the correlators $\langle \zeta \zeta \rangle$ and $\langle \zeta \zeta \zeta \rangle$.
So to compute the evolution of these quantities we need to know the evolution of the correlators of $\delta\varphi_{\alpha}$. The full set of basis vectors contains all information required to determine the evolution of the bundle, encoded in Eq.\eqref{eq:real-deviation} by the $u$-tensors. To obtain transport equations for the correlation functions simply requires reorganisation of this information. As was shown in Refs~\cite{MT1,MT2, optics} this can be done in a number of ways. A particularly quick method is to acknowledge that provided the perturbations can be treated classically, we expect $d \langle O \rangle / d N = \langle d O / d N\rangle$ for any quantity $O$. We can therefore immediately arrive at expressions for the two-point and three-point functions. 
Writing the two-point function as $\Sigma_{\alpha\beta}\equiv \langle \delta \varphi_{\alpha}\delta \varphi_{\beta} \rangle$, Eq.~\eqref{eq:real-deviation} implies
\begin{equation}
\frac{d \Sigma_{\alpha\beta}}{d N}=\left\langle\frac{d \delta\varphi_{\alpha}}{d N}\delta \varphi_{\beta}+\delta \varphi_{\alpha}\frac{d \delta\varphi_{\beta}}{d N}\right\rangle = u_{\alpha\gamma}\Sigma_{\gamma\beta}+ u_{\beta\gamma}\Sigma_{\gamma\alpha}+ \text{[$\geq$ 3 p.f.]}
\label{eq:2pf-transport}
\end{equation}
Similarly, writing the three-point function as $\alpha_{\alpha\beta\gamma}\equiv \langle\delta \varphi_{\alpha}\delta \varphi_{\beta}\delta \varphi_{\gamma} \rangle$, we get
	\begin{equation}
		\frac{d \alpha_{\alpha\beta\gamma}}{d N}
		= 
		u_{\alpha\lambda}
		\alpha_{\lambda\beta\gamma}
		+
		u_{\alpha\lambda\mu}
		\Sigma_{\lambda\beta}
		\Sigma_{\mu\gamma}
		+
		\text{cyclic ($\alpha \rightarrow \beta
		\rightarrow \gamma$)}
		+
		\text{[$\geq$ 4 p.f.]} .
	\label{eq:3pf-transport}
	\end{equation}

This forms a coupled set of ordinary differential equations which can in principle be extended to any n-point correlation function (see Ref.~\cite{Gemma} for an implementation of this technique for the trispectrum). In the context of this work we only care about the power spectrum and bispectrum of $\zeta$, so this set of equations encodes all the information we need for the understanding of superhorizon evolution of our observables.

\subsection{Gauge tranformations}
\label{sec:gauge}

Having seen how to compute the evolution of the field perturbations, now we need to relate these to the primordial curvature perturbation $\zeta$. This can be done by realizing that the curvature perturbation $\zeta$ evaluated at some time $t=t_{c}$ is equivalent on large scales to the perturbation of the number of e-foldings $N(t_{c},t_{*},x)$ from an initial flat hypersurface at $t=t_*$, to a final uniform-density hypersurface at $t=t_c$ \cite{deltaNearly},
\begin{equation}
\zeta(t_c,x) \simeq \delta N(t_c,t_*,x)\equiv N(t_c,t_*,x)-N(t_c,t_*)
\end{equation}
where $N(t_c,t_*) \equiv \int_{*}^{c}H dt$. In the transport method, the hypersurfaces at $t=t_*$ and \mbox{$t=t_c$} are chosen to be infinitesimally separated. 
Expanding $\delta N$ in terms of the initial field perturbations to second order, one obtains
\begin{equation}\label{eq:zeta}
\zeta(t_c,x)=\delta N(t_c,t_*,x)=N,_{\alpha}\delta\varphi_\alpha^* +\frac{1}{2}N,_{\alpha\beta}(\delta\varphi_{\alpha}^*\delta\varphi_{\beta}^*-\langle\delta\varphi_{\alpha}^*\delta\varphi_{\beta}^*\rangle),
\end{equation}
where repeated indices should be summed over, and $N,_{\alpha}$, $N,_{\alpha\beta}$ represent first and second derivatives of the number of e-folds with respect to the fields $\varphi_{\alpha}^*$.\footnote{The subtraction of the correlation function in the second term is due to the fact that this covariance matrix corresponds to the contribution from disconnected diagrams which gives the vacuum energy. In Fourier space one only considers connected diagrams from the outset and thus the subtraction is already implicitly taken care of.} The $N$ derivatives are simply a gauge transformation from field perturbations to curvature perturbations. This gauge transformation only needs to be performed at the time of evaluation of $\zeta$. The fact that it does not need to be transported through superhorizon evolution is a great numerical advantage of this technique.

It is straightforward to express the observables of interest at the time of evaluation within this formalism. The power spectrum is just related to the two-point correlation function of $\zeta$ and can be obtained by \cite{MT1, MT2}
\begin{equation}\label{eq:2p}
P_{\zeta\zeta}=N_{,\alpha}N_{,\beta}\Sigma_{\alpha\beta}.
\end{equation}
To obtain initial conditions at horizon-crossing, we set all the fields to be effectively massless. We tested the consistency of this assumption and found it to be valid in every realisation.\footnote{We would like to thank Sebastien Renaux-Petel for pointing out this possible issue.}
The scalar spectral index, which expresses how the power spectrum changes with scale, is defined as
\begin{equation}
n_{s}-1 \equiv \left. \frac{d\ln{P_{\zeta\zeta}}}{d \ln{k}} \right|_{k = k_*} 
\end{equation}
where $k_*$ is the pivot scale. This can be rewritten as
\begin{equation}
	n_{s}-1 =
		\frac{N_{,\alpha}N_{,\beta}}{P_{\zeta\zeta}}
		\frac{d \Sigma_{\alpha\beta}}{d \ln{k}}
		= \frac{N_{,\alpha}N_{,\beta} n_{\alpha\beta}}
			{N_{,\lambda} N_{,\mu} \Sigma_{\lambda\mu}} ,
	\label{eq:ns}
\end{equation}
where we have introduced the matrix $n_{\alpha\beta} \equiv d \Sigma_{\alpha\beta} / d \ln k$.
Since the gauge-transformation factors $N_{,\alpha}$ and the expansion tensors $u_{\alpha\beta}$
are $k$-independent (they depend only on the
typical trajectory followed by the smoothed fields at each time) we can understand how the spectral index evolves on superhorizon scales \cite{Mafalda}. The only necessary ingredient is a transport equation for the object $n_{\alpha\beta}$ which is:
\begin{equation}
	\label{evolns}
	\frac{d n_{\alpha\beta}}{d N} =
		\frac{d}{d\ln k}
		\frac{d \Sigma_{\alpha\beta}}{d N} =
			u_{\alpha\lambda}n_{\lambda\beta}
			+ u_{\beta\lambda}n_{\lambda\alpha} .
\end{equation}

In this work, for the first time, we apply an equivalent method to evaluate the running of the spectral index, which estimates how $n_{s}$ itself changes with scale.
It is defined as
\begin{equation}
\left.  \frac{d(n_{s}-1) }{d \ln k}  \right|_{k = k_*} = \frac{d}{d\ln k}\left(\frac{N_{,\alpha}N_{,\beta}n_{\alpha\beta}}{P_{\zeta\zeta}} \right)= \frac{N_{,\alpha}N_{,\beta}r_{\alpha\beta}}{P_{\zeta\zeta}}-(n_{s}-1)^2
\end{equation}
where we have introduced the matrix $r_{\alpha\beta} \equiv d n_{\alpha\beta} / d \ln k$. Again, to have the evolution of the running, we just need to specify the transport equation for $r_{\alpha\beta}$ which is:
\begin{equation}
	\label{evolns}
	\frac{d r_{\alpha\beta}}{d N} =
		\frac{d}{d\ln k}
		\frac{d n_{\alpha\beta}}{d N} =
			u_{\alpha\lambda}r_{\lambda\beta}
			+ u_{\beta\lambda}r_{\lambda\alpha} .
\end{equation}
The horizon-crossing initial condition for this expression can be obtained by calculating the derivative with respect to $\ln k$ of the horizon-crossing value of $n_{\alpha\beta}$ given in Ref.~\cite{Mafalda}.\footnote{We would like to thank David Seery for discussions on this topic.} This derivative is to be evaluated at equal times. To do so, we first compute how the $n_{\alpha\beta}$ associated to the pivot scale $k_*$ changes with a variation in scale like $k_*+ \delta \ln k$. This is
\begin{equation}
	\left. \frac{d n_{\alpha\beta}}{d \ln k}\right|_{\rm h.c.}
		= (- 2 \epsilon^{\prime}_{\rm h.c.}+4\epsilon^{2}_{\rm h.c.})
			\Sigma_{\alpha\beta}|_{\rm h.c.} -
			\left. \left(\frac{du_{\alpha\gamma}}{d N}\Sigma_{\gamma\beta}+\frac{du_{\beta\gamma}}{dN}\Sigma_{\gamma\alpha}\right)\right|_{\rm h.c.}+2\epsilon_{\rm h.c.}(u_{\alpha\gamma} \Sigma_{\gamma\beta}+ u_{\beta\gamma}\Sigma_{\gamma\alpha})|_{\rm h.c.}
\end{equation}
where $\epsilon^{\prime}=d\epsilon/dN$.

The $k_*$ and $k_*+ \delta \ln k$ modes cross the horizon at different times but we are looking for the change at \emph{equal} times. When compared at the same time the longer mode experiences slightly more evolution. For this reason, we then need to include an extra contribution that corresponds to this $n_{\alpha\beta}$ displacement.  Remembering that at horizon crossing $d \ln k \sim d N$, this is just $d n_{\alpha\beta}/ dN$.
The total expression is then
\begin{eqnarray}
r_{\alpha\beta}|_{\rm h.c.}=(- 2 \epsilon^{\prime}_{\rm h.c.}+4\epsilon^{2}_{\rm h.c.})
			\Sigma_{\alpha\beta}|_{\rm h.c.} -
			\left. \left(\frac{du_{\alpha\gamma}}{d N}\Sigma_{\gamma\beta}+\frac{du_{\beta\gamma}}{dN}\Sigma_{\gamma\alpha}\right)\right|_{\rm h.c.}\\
+ 2\epsilon_{\rm h.c.}(u_{\alpha\gamma} \Sigma_{\gamma\beta}+ u_{\beta\gamma}\Sigma_{\gamma\alpha})|_{\rm h.c.}-(u_{\alpha\gamma}n_{\gamma\beta}+u_{\beta\gamma}n_{\gamma\alpha})|_{\rm h.c.}.\nonumber
\end{eqnarray}

The local non-gaussianity parameter $f_{\rm NL}$ is defined as \cite{deltaN}
\begin{equation}
f_{\rm NL} \equiv \frac{5}{18}\frac{B_{\zeta\zeta\zeta}}{(P_{\zeta\zeta})^2}
\end{equation}
where $B_{\zeta\zeta\zeta}$ is the bispectrum of curvature perturbations, related to the three-point function of $\zeta$. It is useful to decompose it as $B_{\zeta\zeta\zeta}=B_{\zeta\zeta\zeta1}+B_{\zeta\zeta\zeta2}$,
where 
\begin{equation}\label{eq:3p1}
B_{\zeta\zeta\zeta1}=N,_{\alpha}N,_{\beta}N,_{\gamma}\alpha_{\alpha\beta\gamma},
\end{equation} 
and
\begin{equation}\label{eq:3p2}
B_{\zeta\zeta\zeta2}=\frac{3}{2}N,_{\alpha}N,_{\beta}N,_{\gamma\rho}\left[\Sigma_{\alpha\gamma}\Sigma_{\beta\rho}+\Sigma_{\alpha\rho}\Sigma_{\beta\gamma}\right].
\end{equation} 
Eq.~\eqref{eq:3p1} is the intrinsic non-linearity among the fields, while Eq.~\eqref{eq:3p2} encodes the non-Gaussianity resulting from the gauge transformation to $\zeta$ \cite{MT1,MT2}.\\

Given these expressions for the observables, it is now only necessary to specify the gauge transformation expressed by the $N$ derivatives. Following the procedure developed in Ref.~\cite{us}, the expansion of Eq.~\eqref{eq:zeta}, can be understood in two steps. First, $\delta N$ can be written as an expansion in terms of $\delta \rho$, where $\delta$ refers to a change from an initial flat hypersurface to a final uniform-density hypersurface:\footnote{We particularly thank David Mulryne for this result and discussions around this topic.}
\begin{equation}
\delta N=\frac{dN}{d\rho}\delta\rho+\frac{1}{2}\frac{d^2N}{d\rho^2}\delta\rho^2+ \cdots,
\end{equation}
where $\rho=3H^{2}$. To obtain the derivatives of $N$ as desired, one just needs to perturb each term of the above expansion in terms of the fields. The result is \cite{us}
\begin{equation}
N_{,\phi_{i}}=\frac{V_{,i}}{2H^2\epsilon(3-\epsilon)},
\end{equation}
\begin{equation}
N_{,p_{i}}=\frac{VTp_{i}}{2H^2\epsilon(3-\epsilon)^2},
\end{equation}
\begin{equation}
N_{,\phi_i\phi_j}=\frac{V_{,ij}}{2H^{2}\epsilon(3-\epsilon)}-\frac{V_{,i}V_{,j}}{4H^4\epsilon^2(3-\epsilon)^2}\left(\frac{Tp_{k}{p}^{\prime k}}{\epsilon}
+2\epsilon\right),
\end{equation}
\begin{equation}
N_{,p_ip_j}=\frac{T\delta_{ij}}{2\epsilon(3-\epsilon)}-\frac{T^2p_{i}p_{j}}{4\epsilon^2(3-\epsilon)^2}\left(\frac{Tp_{k}{p}^{\prime k}}{\epsilon}
-6\epsilon+12\right),
\end{equation}
\begin{equation}
N_{,\phi_ip_j}=-\frac{TV_{i}p_{j}}{4H^2\epsilon^2(3-\epsilon)^2}\left(\frac{Tp_{k}{p}^{\prime k}}{\epsilon}
-2\epsilon +6\right),
\end{equation}
where the derivatives are taken explicitly with respect to $\phi_{i}$ and $p_{i}$ rather than the combined $\varphi_{\alpha}$, the objects $V_{,i}$ and $V_{,ij}$ refer to the derivatives of the potential with respect to the fields $\phi_{i}$ and $\phi_{j}$ and 
$p^\prime_{i}=dp_{i}/dN=d^2\phi_{i}/d^2N$.

\subsection{The adiabatic limit}

This section has been dedicated to the understanding of superhorizon evolution of the curvature perturbations. What is yet to be addressed is at what point $\zeta$ ceases to evolve. This happens when the trajectory has effectively reached its adiabatic limit, \textit{i.e.}\ the model has become effectively single-field (see Ref.~\cite{ALearly} for early discussions on this topic).

One needs to ensure that the adiabatic limit is reached before the time of computation of observables, which generally is taken to be at the end of inflation. If isocurvature modes are still present at that point, the curvature perturbations will continue to evolve through an epoch of reheating, for which there is no precise knowledge. This would mean that the conclusions reached at the time of estimation would necessarily be incomplete, and the model not predictive.

In the D-brane case, the situation is even more delicate, since the 
validity of the approximations of the set-up breaks down as one approaches the tip of the conifold, or, for practical purposes, at $x=0.02$.
If $\zeta$ continues to evolve after that point, it will suffer changes due not only to reheating, but also an unknown background geometry and potential.

Formally, the approach to an adiabatic limit can be understood via the parameter $\theta={\rm tr} \ u_{\alpha\beta}$ of Eq.~(\ref{eq:expansion-tensor}) which describes the tendency for the bundle of trajectories to focus or dilate. The factor by which  the  bundle's cross-section has grown (or decayed) from its horizon-crossing value at $N_{0}$ to the time $N$ is given by:
\begin{equation}\label{eq:thetah}
\Theta(N,N_{0}) \equiv \exp \left[ \int^{N}_{N_{0}}\theta(n)dn \right].
\end{equation}
In a slow-roll two-field case, where the phase space is simply $\phi_{i}$, when $\Theta \rightarrow 0$, or equivalently $\theta \rightarrow -\infty $, the trajectory is becoming effectively single-field \cite{optics}. However, since we are working with more than two fields and in a phase space composed by $\phi_{i}$ and $ p_{i}$ the situation is more subtle, and the relation not so direct \cite{optics, us}. 

In the full phase space, it is instructive to understand how the bundles of $\phi_{i}$ and $p_{i}$ individually behave. In particular, let us identify $p_{i}=p_{i}^{\rm SR}(\phi_{i})+s_{i}$, where $p_{i}^{\rm SR}$ is the slow-roll attractor for the momenta $p_{i}$ and a function of $\phi_{i}$ only and $s_{i}$ are the momenta isocurvature modes.
With these new variables, we can write \cite{us}
\begin{equation}
\theta=\theta^{\rm SR}(\phi_{i})+\theta^{s}(s_{i})
\end{equation}
where $\theta^{\rm SR}$ describes the dilation of the field bundle and $\theta^{s}$ describes the dilation of the $s$ modes, \textit{i.e.}\ how the momenta converge to their slow-roll attractor. We can also define $\Theta^{\rm SR}$ and $\Theta^{s}$ in analogy with Eq.~(\ref{eq:thetah}).

The interpretation of when the adiabatic limit has been adequately reached from the above quantities is not straightforward \cite{us}, but to infer when it has not been is often simple. For the purpose of this work it will be sufficient to know that if $\Theta^{\rm SR} \gtrsim 1$ then an adiabatic limit certainly has not been reached. 

\section{Distributions for observables}

We now present the results obtained from computing the curvature perturbations for different samples of inflationary trajectories. The outcomes are distributions for the values of the cosmological parameters presented in the previous section: amplitude $P_{\zeta\zeta}$, spectral index $n_{s}$ and running of the spectral index of the scalar power spectrum, tensor-to-scalar ratio $r$, and local non-gaussianity parameter $f_{\rm NL}$. We compare these with constraints from observations; all constraint contours are $95\%$ confidence limits using the WMAP 7 year data release combined with baryonic acoustic oscillations and supernov$\ae$  data \cite{WMAP7,WMAP5,marina}. We use as pivot the scale that crossed the horizon 55 e-folds before the end of inflation. 

\subsection{Field number dependence}

\begin{figure}[t]
\centering
\includegraphics[width=15cm]{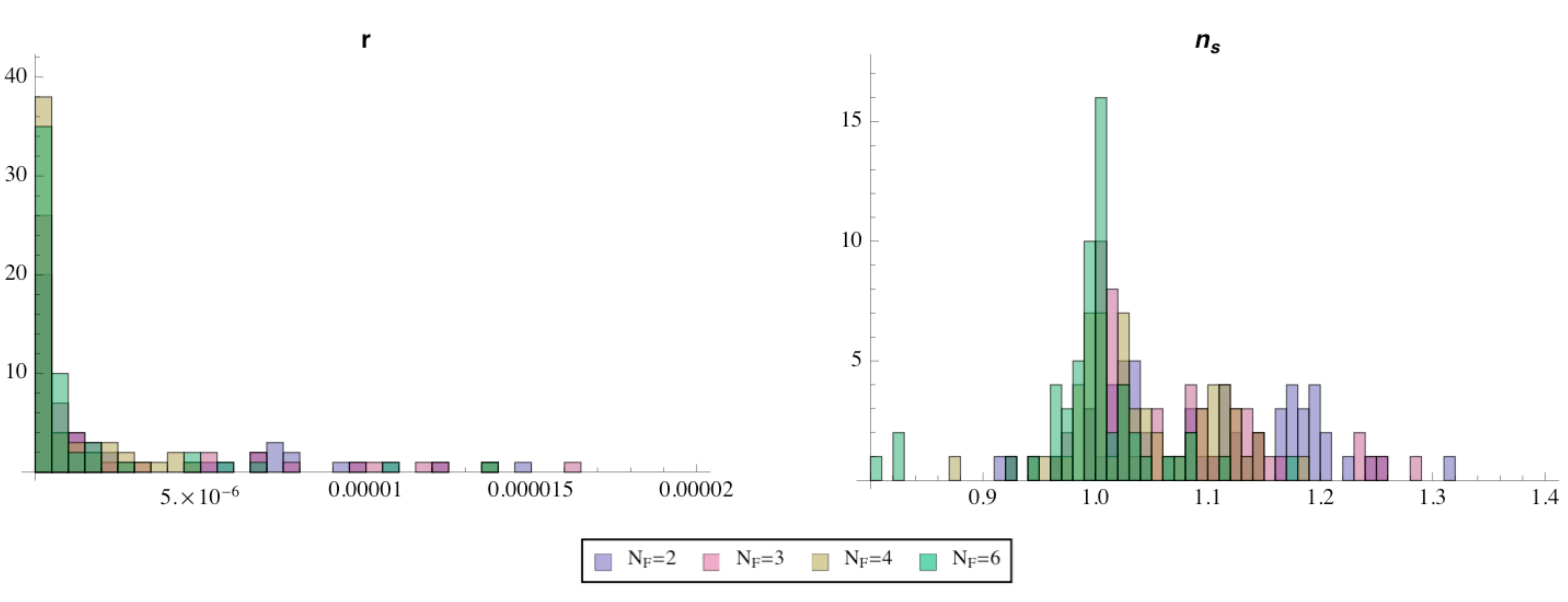}
\caption{Distributions for the tensor-to-scalar ratio $r$, left, and scalar spectral index $n_{s}$, right. The distributions were taken for different numbers of active fields. It is possible to identify a small suppression in $r$ and a tendency towards redder values of $n_{s}$ (keeping the peak at $n_{s}=1$) as $N_{\rm F}$ increases.}
\label{dtest}
\end{figure}

\begin{figure}[t]
\centering
\includegraphics[width=8cm]{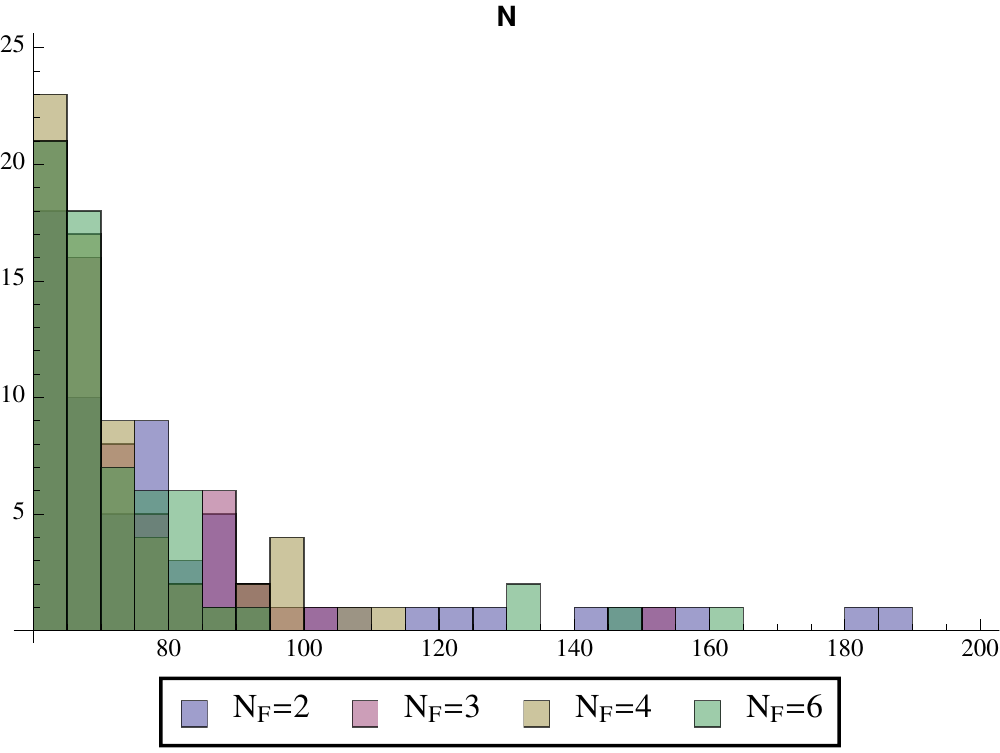}
\caption{Number of e-folds of inflation for different number of active fields. We can see that this distribution is unchanged by varying $N_{\rm F}$.}
\label{Nends}
\end{figure}
Regarding the dependence of the inflationary behaviour on the number of active fields, we found that the effect of increasing the number of angular directions is negligible for the trajectories and consequent observables. Fig.~\ref{dtest} shows how the increase in angular directions changes the distributions of the tensor-to-scalar ratio and spectral index. We can identify a slight suppression in $r$ as the number of active fields increases, as one would expect since more directions allow for more turns in field space which fuel the scalar power spectrum. The spectral index, which consistently peaks at 1, gets a shift towards red values as the number of fields increases. This is not the consequence of a change in the inflationary trajectories and it can be readily understood, as will be shown shortly.

Interestingly, although an increase in the number of fields allows in principle a wider range of trajectories, we did not encounter any difference in the distribution for the total number of e-folds when varying $N_{\rm F}$, as can be seen in Fig.~\ref{Nends}.

In the remainder of this paper, we will concentrate mainly on the results for two active fields for which our sample is constituted of 1086 inflationary trajectories, using the sample for the full six-field case, composed of 93 trajectories, for comparison.

\begin{figure}[t]
\centering
\includegraphics[width=10cm]{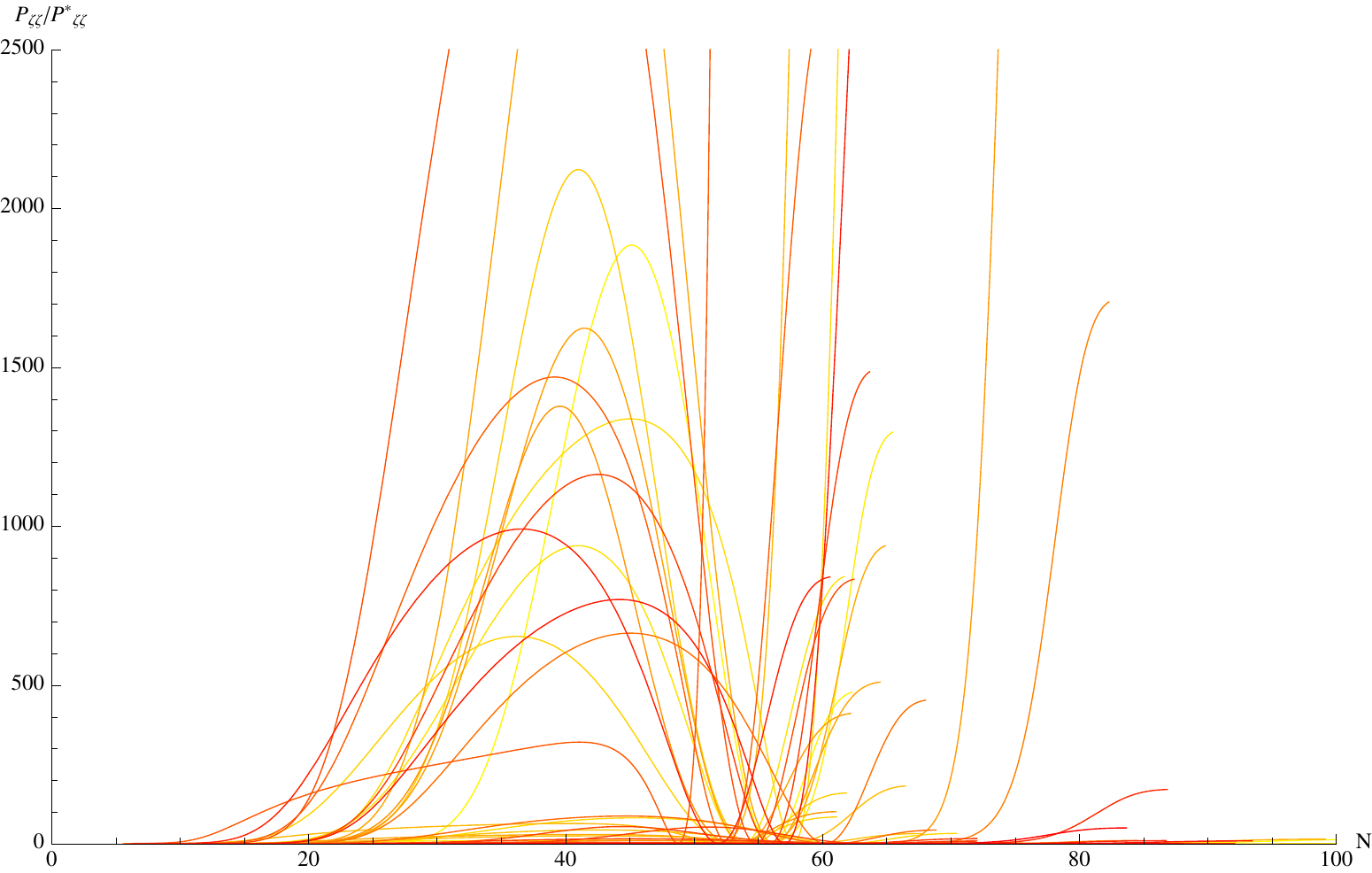}
\caption{Superhorizon evolution of the amplitude of the scalar power spectrum $P_{\zeta\zeta}$ for 50 trajectories with 2 active fields, coloured at random. The evolutions show a consistent non-monotonic growth that we refer as `caterpillar' shaped. It is impractical to show the full evolution of all the trajectories due to the large range in scales even for this reduced sample; for this reason the plot has been cut-off.}
\label{evolutions}
\end{figure}

\subsection{Observables}

The amplitude of the power spectrum, as can be seen in Fig.~\ref{evolutions}, consistently undergoes a superhorizon evolution that we refer to as `caterpillar' shaped, \textit{i.e.}\ non-monotonic. This is interesting as this sort of evolution is very rare in purely random generated potentials~\cite{me2} and so indicative of a dynamical trait common to all of our inflation realisations. The next section will explore its origin. As can be seen in Fig.~\ref{hists}, the histogram of $P_{\zeta\zeta}$ has a smooth maximum at around $10^{-9}$, in agreement with observations (the WMAP value is $\sim 2.5 \times 10^{-9}$~\cite{WMAP5}). This is not surprising as the overall magnitude of the potential is determined by the scale $\mu^{4}$, which in turn is set by our choice of the throat length $r_{\rm UV}$. The important fact is that the distribution does not sharply peak at a precise value of $P_{\zeta\zeta}$, indicating that there is no fine-tuning issue around this parameter. 

The spectral index shows a much clearer peak at $n_{s}=1$, as seen in Fig.~\ref{hists}. Actually, the spread in  $n_{s}$ is not well approximated by a Gaussian; instead, two different populations can be identified, one with $n_{s} \geq1$ and one with $n_{s}<1$. 
For the two-field ensemble, these correspond to $\sim 84\%$ and $\sim 16\%$ of the trajectories, respectively. As seen previously, an increase in the number of active fields enhances the number of red trajectories; the respective ratios, for the six-field sample are then $\sim 50\%$ each. 
In the next section we will address the dynamical characteristics of these two populations and the changes with the number of fields.

\begin{figure}[t]
\centering
\includegraphics[width=16cm]{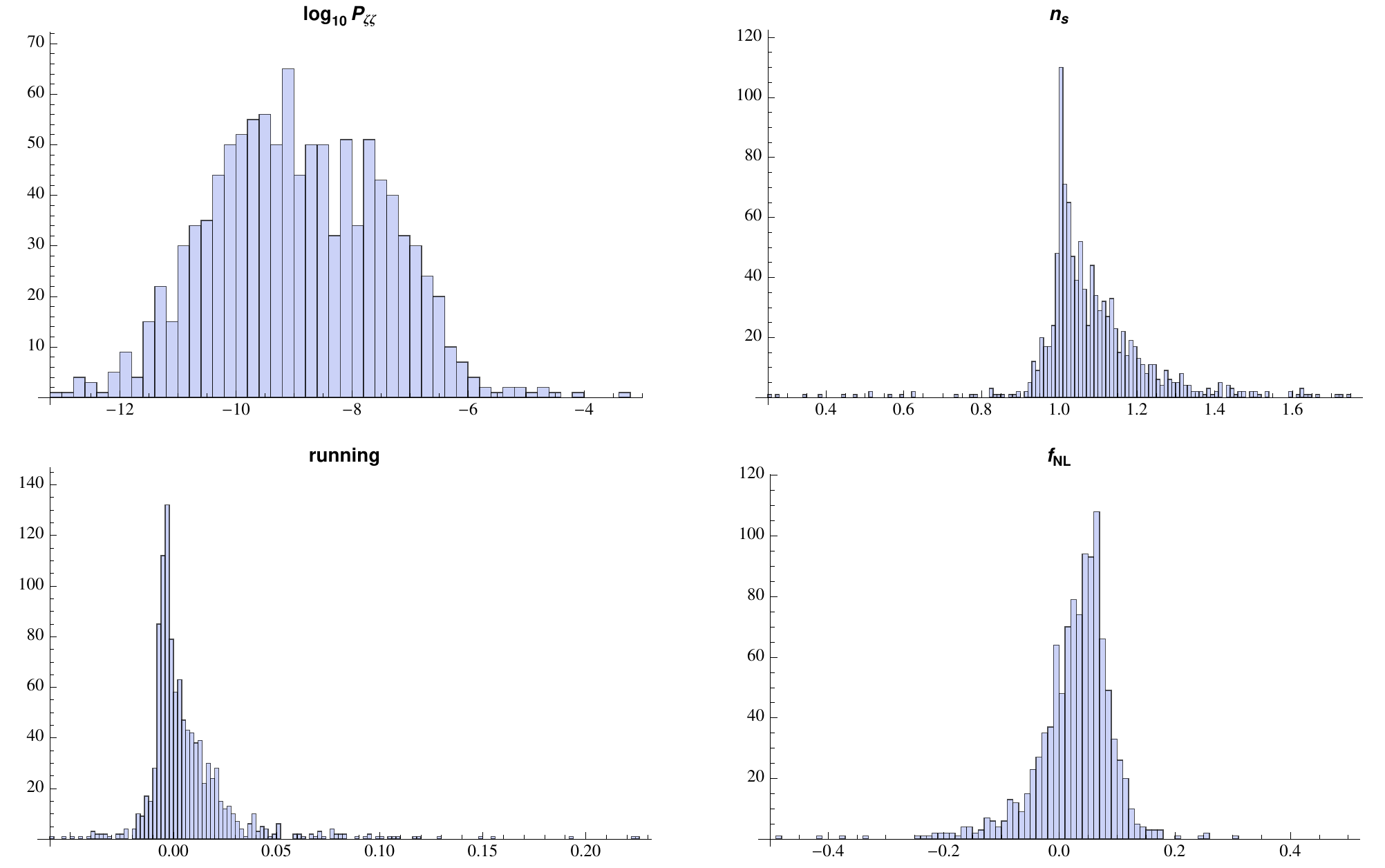}
\caption{Distributions for the amplitude of the power spectrum $P_{\zeta\zeta}$, top left, scalar spectral index $n_{s}$, top right, running of the spectral index, bottom left, and local non-gaussianity parameter $f_{\rm NL}$, bottom right. All distributions were taken for the sample with 2 active fields.}
\label{hists}
\end{figure}

A very interesting result comes from the computation of the running of the spectral index. Fig.~\ref{hists} shows that the running tends to be positive and that it can take large values. This outcome, as will be discussed shortly, is extremely constraining. One could think that, since the running can take large values, the value of the spectral index changes a lot with the choice of pivot scale. Our approach is to think that sampling over a large number of inflationary potentials reproduces the effect of sampling over different choices of pivot scales, such that the final distributions for different pivot scales are actually identical. We tested this assumption and found it to be the case. 

\begin{figure}[t]
\centering
\includegraphics[width=15cm]{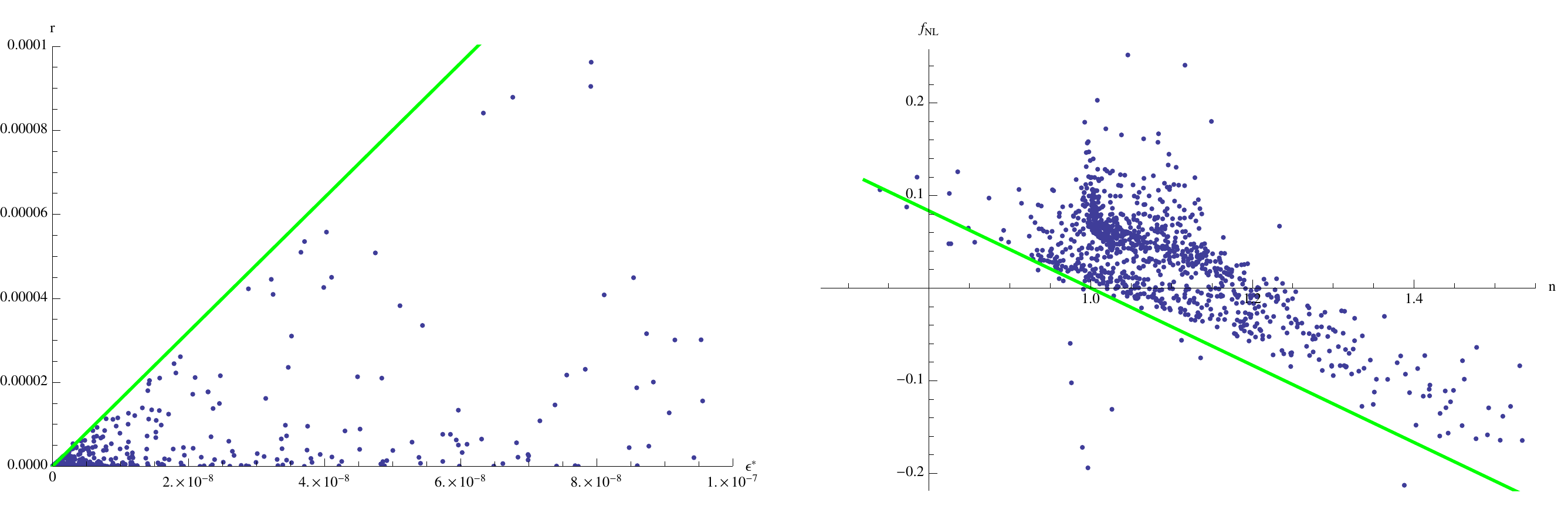}
\caption{Plot of the values of $\epsilon$ at horizon-crossing versus  $r$, left, and $n_{s}$ versus $f_{\rm NL}$, right, for the sample with 2 active fields. The green lines represent the single-field predictions, $r=16\epsilon/T_{3}$ and  $f_{\rm NL}=-5(n_{s}-1)/12$, respectively.}
\label{scats1}
\end{figure}

The tensor-to-scalar ratio is always extremely small, as it is related to the slow-roll parameter $\epsilon$ that remains $\ll 1$ throughout the calculation. This can be clearly seen in Fig.~\ref{scats1}. Furthermore, the usual single-field result for the brane case would be the relation
\begin{equation}
r=16\frac{\epsilon}{T_{3}}
\end{equation}
which corresponds to the green line. We can see how multifield effects change this result, by weakening this expression to an inequality. 

The local non-gaussianity parameter $f_{\rm NL}$, as can be seen in Fig.~\ref{hists}, is almost always too small to possibly be detected by any anticipated observation. In the full two-field ensemble only five trajectories yielded values of $|f_{\rm NL}| > 1$. Although these are highly unlikely, any case presenting interesting observational signatures can be informative in its own right; we leave a detailed analysis of these cases for future work.
When plotted against $n_{s}$, this parameter also shows the deviation from the single-field prediction  $f_{\rm NL}=-5(n_{s}-1)/12$~\cite{Maldy} represented by the green line in Fig.~\ref{scats1}. Once again, multifield effects break this degeneracy. \\

\subsection{Constraints from WMAP}

We now impose the observational constraints on the distributions.

As can be seen in Fig.~\ref{scats2}, the majority of the trajectories gives rise to values of running and spectral index that lie outside the observational bounds. From the running versus $n_{s}$ plot alone, one could think that the trajectories giving rise to the peak around $n_{s}=1$ would be in agreement with observations. Actually, a stronger constraint for $n_{s}$ is imposed by the fact that the model predicts a negligible tensor-to-scalar ratio. Constraints on the $n_{s}$ versus $r$ plot shown in Fig.~\ref{scats2} exclude all the trajectories which result in $n_{s} > 0.995$. 
These constraints alone result in only $\sim 10\%$ of the total two-field sample, and $\sim 50\%$ of the six-field sample, of trajectories being in agreement with observations.

\begin{figure}[t]
\centering
\includegraphics[width=15cm]{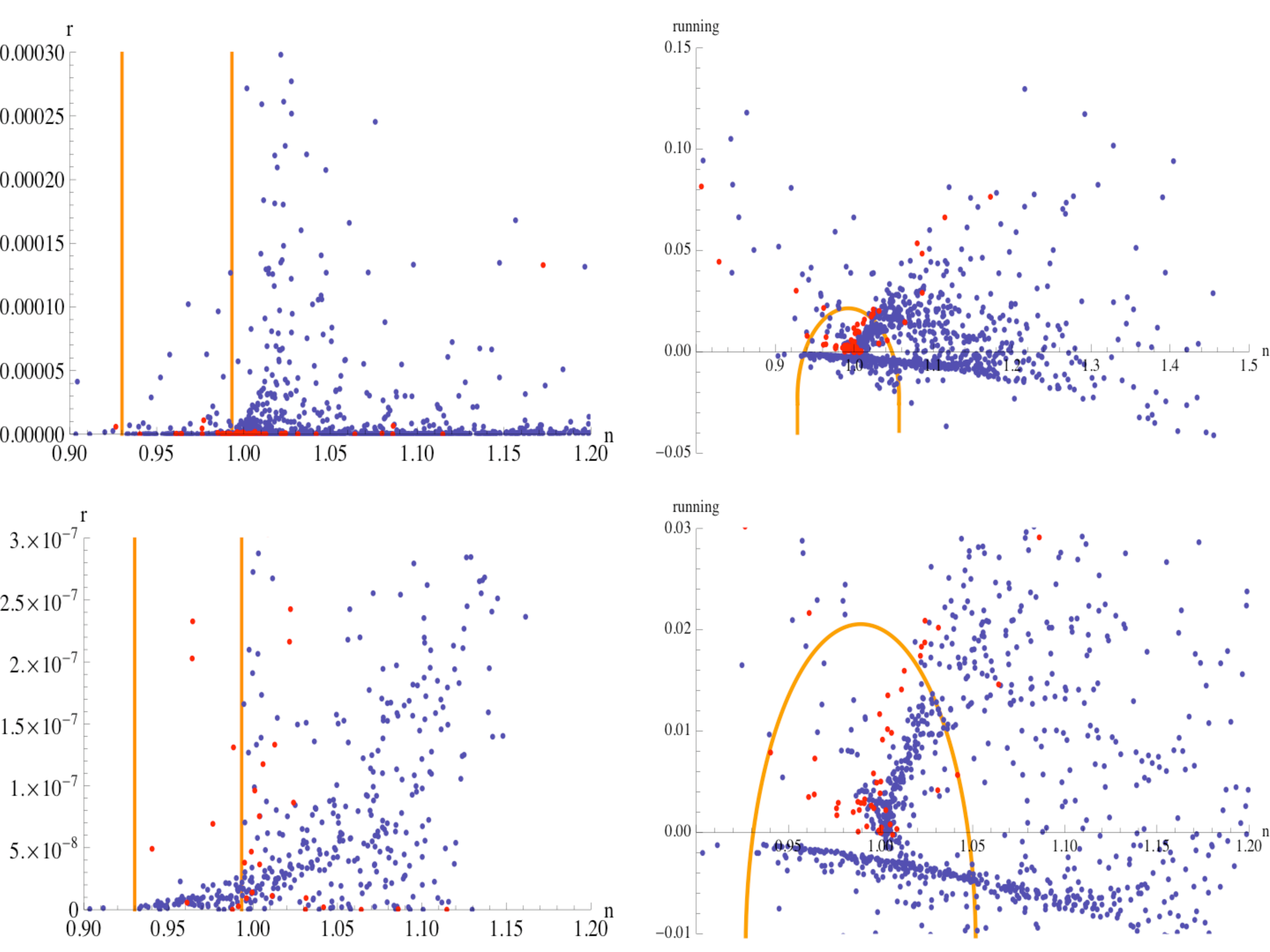}
\caption{Plot of the values of $n_{s}$ versus  $r$, left, and $n_{s}$ versus running, right, for the model with 2 active fields, blue points, and 6 active fields, red points. The lower panels greatly expand the vertical scales. The orange lines represent the $95\%$ confidence limits using WMAP data. When running is allowed to be $\neq 0$ it is important to know what is the best scale with which to make comparisons with observations. In Ref.~\cite{marina} this issue was explored; following their conclusions, we look at the constraints for $k^{*}=0.017 \, {\rm Mpc}^{-1}$ which is not the same choice made in Ref.~\cite{WMAP5}.} 
\label{scats2}
\end{figure}

A further constraint is imposed by requiring the correct amplitude of the scalar power spectrum, $(2.5 \pm 0.1) \times 10^{-9}$ \cite{WMAP5}. Combining all constraints we obtained only two realisations in total concordance with observations in the full sample of 1086 cases of two-field inflation, and one realisation in the sample of 93 cases of six-field inflation. As discussed in the previous subsection, this is not a worrying result as the distribution of $P_{\zeta\zeta}$ does not show a sharp peak.

\section{A close look at trajectories}

In this section we describe in detail the dynamics of individual trajectories and how they give rise to the distributions seen in the previous section. 

A remarkable feature of this model is that all inflationary trajectories encountered were essentially of the same type, inflection-point inflation with a wiggle in the trajectory. It turns out that while in principle inflation could occur at any location within the throat (above the tip), at least the last 60 e-folds of inflation always occur in a small sub-region (typically $0.02< x< 0.09$) in the vicinity of the inflection point discussed in~\S\S\ref{sec:D-brane potential}. While the inflection-point contribution to the potential, $V_{C} + V_{M}$, alone could not give enough inflation, the contributions from the bulk  
can alter the potential in such a way that sufficient inflation \emph{can} occur. The range of contributions capable of giving inflation is limited so in practice the result is a very consistent dynamical behaviour.


\subsection{One inflationary trajectory to explain them all}

\begin{figure}[t]
\centering
\includegraphics[width=15cm]{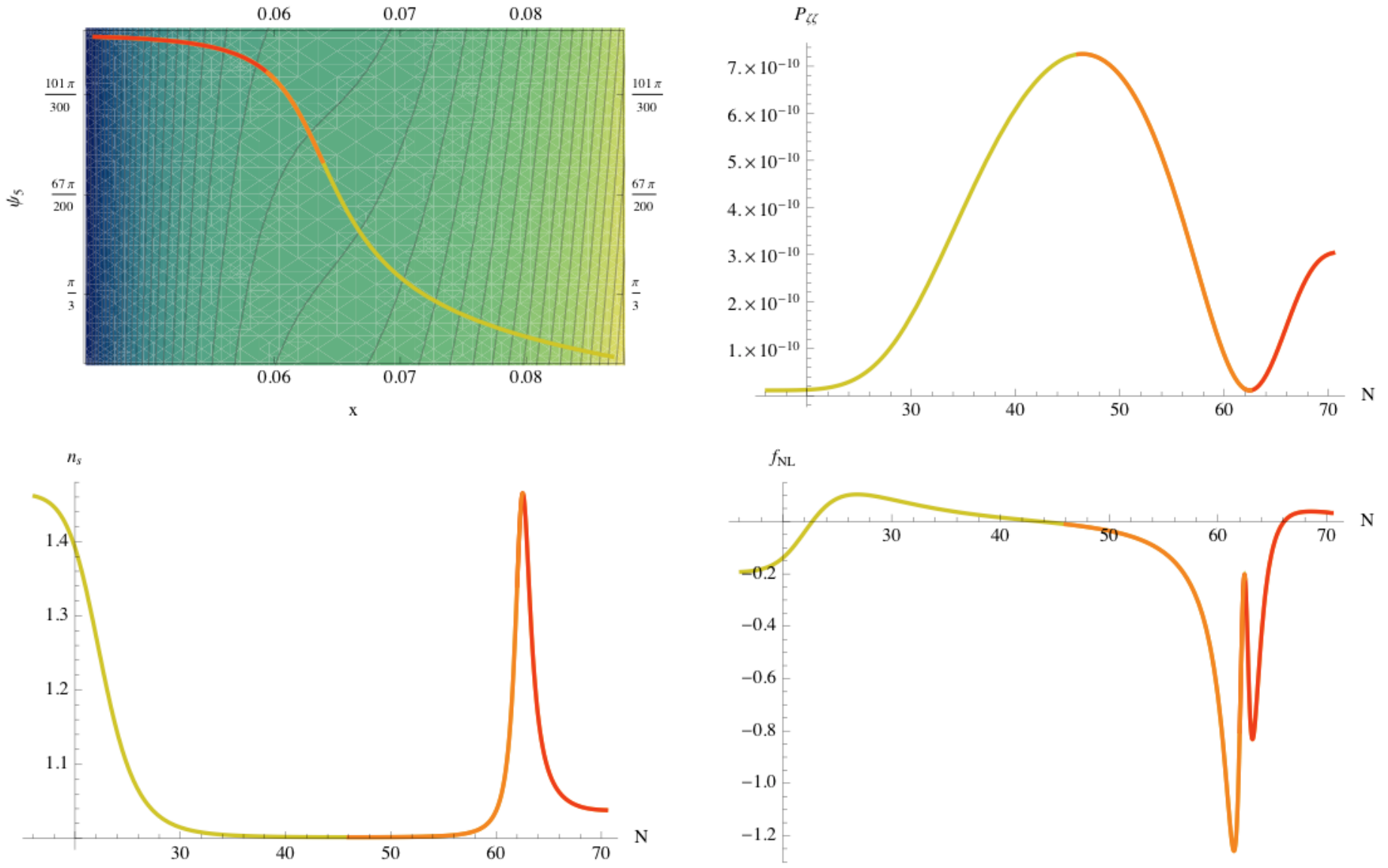}
\caption{The dynamical behaviour of Verse 20277, for the region of the trajectory where inflation occurred. The top left plot shows the trajectory in field space plotted on top of the potential contours. The evolution of the amplitude of the power spectrum, top right, spectral index, bottom left, and running, bottom right, show different colours regarding the position in the potential; yellow represents evolution before the inflection point, orange between the inflection point and the fall of the ledge and red after the starting falling off the ledge. For the region represented in red, $|\eta|>1$.}
\label{fig:trajetc}
\end{figure}

A typical inflationary realisation in our $N_{\rm F}=2$ set-up is the affectionately-named Verse 20277. Fig.~\ref{fig:trajetc} shows the inflationary trajectory superimposed on a contour plot of the potential, together with the evolution of the observable quantities we investigated. The trajectory evolves from right to left, passing an inflection point in the radial direction but also experiencing a slope in the angular direction, causing a wiggle in what would otherwise be standard single-field inflection-point inflation (see for example Ref.~\cite{inflection}).  As will be shown, this behaviour explains the non-monotonic evolution of $P_{\zeta\zeta}$, which we saw to be so common. Verse 20277, as $\sim 84\%$ of our realisations, has a blue spectral index, $n_{s}>1$, which can also be understood by its inflationary dynamics.  It also presents the most common evolution of $f_{\rm{NL}}$ we encountered, giving rise to an unobservably small value of $f_{\rm{NL}}$, as seen consistently across our distributions.

Another interesting characteristic of the dynamics of Verse 20277, which again is representative of the whole sample, is that the process of falling off the ledge after the inflection point gives rise to a prolonged period of non-slow-roll inflation, often leading to ${\cal O}(10)$ e-folds. This regime is shown in red in Fig.~\ref{fig:trajetc}. While the slow-roll parameter $\epsilon$ remains much smaller than 1 throughout all the analysis, the parameter $|\eta|$ increases significantly, getting to values of $\sim 35$. The consequences of this effect will be discussed in what follows.


\subsection{A separable potential approximation}

A simple explanation for all of the above evolutionary traits presents itself if we approximate the inflationary region as a separable potential of the type.\footnote{Approximating small regions of potentials as separable is a very useful technique which can be applied quite generally. See Ref.~\cite{JDD} for another recent and rather nice example of this.} 
\begin{equation}
W(\phi)\equiv W(x,\Psi)= U(x)+\sum_{i=2}^{N_{\rm F}}V_{i}(\Psi_{i})
\end{equation}
where $U(x)$ possesses an inflection point in the radial direction
\begin{equation}
U(x)\equiv V_{1}(\phi_{1})=\alpha_{0}+\alpha_{1}(x-x_{0})+\alpha_{3}(x-x_{0})^3
\end{equation}
and $V(\Psi)$ are slopes in the angular directions
\begin{equation}
V_{i}(\Psi_{i})= \beta_{i} \Psi_{i}.
\end{equation}
As already mentioned, for the majority of inflation, the evolution sits well within the slow-roll regime. During this period, we can write the number of e-folds of inflation as 
\begin{equation}
N(t_{c},t_{*})=-T_{3}\int_{*}^{c}\sum_{i=1}^{N_{\rm F}}\frac{V_{i}}{V_{,i}}d\phi_{i}.
\end{equation}
For this section we take the flat surface to be at horizon-crossing and the constant density surface to be at our time of evaluation.
This allows us to write down the total derivative as
\begin{equation}
dN=T_{3}\sum_{j=1}^{N_{\rm F}}\left[\left(\frac{V_{j}}{V_{,j}}\right)-\sum_{i=1}^{N_{\rm F}}\frac{\partial \phi^{c}_{i}}{\partial \phi^{*}_{j}}\left(\frac{V_{i}}{V_{,i}}\right)\right]d\phi_{j}^{*}.
\end{equation}
Following the procedure in Ref.~\cite{sep} one can obtain an expression for the partial derivative in the previous expression as
\begin{equation}
\frac{\partial\phi^{c}_{i}}{\partial\phi_{j}^{*}}=-\frac{W_{c}}{W_{*}}\sqrt{\frac{\epsilon^{c}_{i}}{\epsilon^{*}_{j}}}\left(\frac{\epsilon^{c}_{j}}{\epsilon^{c}}-\delta_{ij}\right)
\end{equation}
where the slow-roll parameters associated with each field are defined as
\begin{equation}
\epsilon_{i}\equiv\frac{1}{2T_{3}}\left(\frac{V_{,i}}{W}\right)^2,\quad \eta_{i}\equiv\frac{1}{T_{3}}\frac{V_{,ii}}{W},\quad  \xi_{i} \equiv \frac{V_{,i}V_{,iii}}{T^2_{3}W^{2}}
\end{equation}
such that $\epsilon=\sum\epsilon_{i}$. Finally, we arrive at an expression for the derivatives of $N$
\begin{equation}
\frac{\partial N}{\partial\phi_{i}^{*}}=\sqrt{\frac{T_{3}}{2\epsilon^{*}_{i}}}\frac{V_{i}^{*}+Z_{i}}{W^{*}},
\end{equation}
where we have the, rather important for the following discussion, term
\begin{equation}\label{eq:Z}
Z_{i}\equiv \frac{1}{\epsilon^{c}}\sum^{N_{\rm F}}_{j=1}V_{j}^{c}(\epsilon_{i}^{c}-\epsilon^{c}\delta_{ij})
\end{equation}
which contains all the information about the constant density surface at the time of evaluation. All other terms are determined by horizon-crossing values.
Note that due to assuming slow-roll, we have reduced our phase space to the $N_{\rm F}$ fields $\phi_{i}$ and  that by approximating the potential as sum-separable, this expression no longer requires the `c' and `*' surfaces to be infinitesimally separated. 

The second derivatives are given by \cite{sep}
\begin{equation}
\frac{\partial^2 N}{\partial\phi^{*}_{i}\partial\phi^{*}_{j}}=\delta_{ij}T_{3}\left(1-\frac{\eta^{*}_{j}}{2\epsilon^{*}}\frac{V^{*}_{j}+Z_{j}}{W_{*}}\right)+\frac{1}{W_{*}}\sqrt{\frac{T_{3}}{2\epsilon^{*}_{j}}}\frac{\partial Z_{j}}{\partial\phi_{i}^{*}}.
\end{equation}

Having obtained expressions for the first and second derivatives of $N$, it is now possible to write down analytical expressions for the desired observable quantities. This is done by using the expressions from~\S\S\ref{sec:gauge} and using the horizon-crossing values for the correlation functions of $\delta \phi$. The amplitude of the power spectrum is given by 
\begin{equation}
\label{Pzeta}
P_{\zeta\zeta}=\frac{T_{3} W_{*}}{24 \pi^2}\sum_{i=1}^{N_{\rm F}}\frac{u_{i}^2}{\epsilon^{*}_{i}}
\end{equation}
where 
\begin{equation}
u_{i}\equiv\frac{V^{*}_{i}+Z_{i}}{W_{*}}.
\end{equation}
The spectral index is given by  \footnote{Here we have used the fact that $d/d \ln k\approx d/d N=(\phi_{i}/H) \partial/\partial \phi_{i}$ and continually made use of the substitutions $N_{,i}V_{,i}=-VT_{3}$ and $\dot{\phi}_{i}N_{,ij}/H=V_{,j}/V+V_{,ij}N_{,i}/VT_{3}$.} 
\begin{equation}\label{eq:n}
n_{s}-1=-2\epsilon_{*}-4\frac{\left(1-\sum_{i=1}^{N_{\rm F}}\frac{\eta^{*}_{i}u_{i}^2}{2\epsilon^{*}_{i}}\right)}{\sum_{i=1}^{N_{\rm F}}\frac{u_{i}^2}{\epsilon^{*}_{i}}}
\end{equation}
Differentiating with respect to $\ln k$ we also obtain a new expression for the running. We find it to be
\begin{equation}\label{eq:runsr}
n^{\prime}=-8\epsilon^{*2}
+4\sum_{i=1}^{N_{\rm F}} \epsilon^{*}_{i}\eta^{*}_{i} 
-16\frac{\left(1-\sum_{i}\frac{\eta^{*}_{i}u_{i}^2}{2\epsilon^{*}_{i}}\right)^2}{\left(\sum_{i}\frac{u_{i}^2}{\epsilon^{*}_{i}}\right)^2}
-8\frac{\sum_{i}\eta^{*}_{i}u_{i}\left(1-\frac{\eta^{*}_{i}u_{i}}{2\epsilon^{*}_{i}}\right)}{\sum_{i}\frac{u_{i}^2}{\epsilon^{*}_{i}}} 
+4\epsilon^{*}\frac{\sum_{i}\frac{\eta^{*}_{i}u^{2}_{i}}{\epsilon^{*}_{i}}}{\sum_{i}\frac{u_{i}^2}{\epsilon^{*}_{i}}}-2\frac{\sum_{i}\frac{\xi^{*}_{i}u^2_{i}}{\epsilon^{*}_{i}}}{\sum_{i}\frac{u_{i}^2}{\epsilon^{*}_{i}}}
\end{equation}
Finally, the local non-Gaussianity is
\begin{equation}\label{eq:fnlsr}
-\frac {5}{6}f_{\rm NL}=2\frac{\sum_{i=1}^{N_{\rm F}}\frac{u_{i}^{2}}{\epsilon_{i}^{*}}\left(1-\frac{\eta_{i}^{*}}{2\epsilon^{*}_{i}}\right)+\sum_{i,j=1}^{N_{\rm F}}\frac{u_{i}u_{j}}{\epsilon^{*}_{i}\epsilon^{*}_{j}}A_{ji}}{\left(\sum_{i=1}^{N_{\rm F}}\frac{u_{i}^2}{\epsilon^{*}_{i}}\right)^2},
\end{equation}
where as a result of differentiating $Z_{i}$, another term $A_{ji}$ containing contributions from the  `c' surface is required \cite{sep};
\begin{eqnarray}
\frac{\partial Z_j^c}{\partial\phi_i^*}&=&-\frac{W^2_c}{W_*}\sqrt{\frac{2}{\epsilon_i^*}}\left[\sum_{k=1}^{N_{\rm F}}\epsilon_k\left(\frac{\epsilon_j}{\epsilon}-\delta_{jk}\right)\left(\frac{\epsilon_i}{\epsilon}-\delta_{ik}\right)\left(1-\frac{\eta_k}{\epsilon}\right)\right]_c \label{finalZ}\\
&\equiv&\sqrt{\frac{2}{\epsilon_i^*}}W_*\mathcal{A}_{ji}\,. \nonumber\label{defA}
\end{eqnarray}
All of the above expressions can be shown to reduce to the standard single-field formula by setting $u_{i}=1$.

With these expressions from the separable potential approximation, the phenomenology described at the beginning of this section becomes clear. We now discuss each of the trends encountered in the results individually. 

\subsection{Why so many caterpillars?}

First, we would like to address the question of why non-monotonic evolution in the amplitude of the power spectrum is so common. 

\begin{figure}[t]
\centering
\includegraphics[width=15cm]{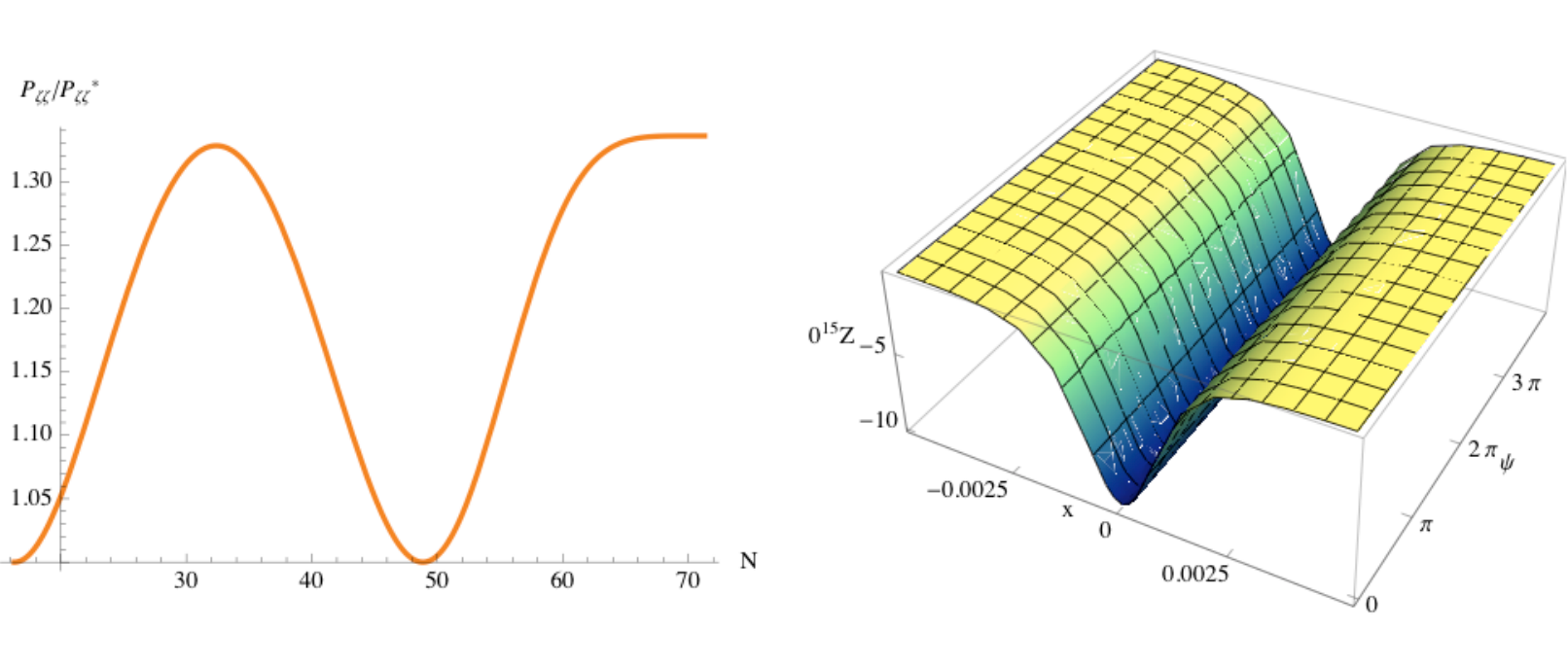}
\caption{Example of evolution of the power spectrum $P_{\zeta\zeta}$ and the form of $Z_{i}$ giving rise to it for the separable potential. The power spectrum appears to stop evolving in the last few e-folds despite the bundle dilating in this region.}
\label{fig:Z}
\end{figure}

Under the separable approximation, all superhorizon behaviour is encapsulated in the $Z_{i}$ terms. This must go to zero in the adiabatic limit since, as there is no evolution in this limit, the result must be independent of our choice of the time of evaluation. However this is not a sufficient condition for reaching the adiabatic limit; as we will see, it is possible for  $Z_{i}$ to become negligible towards the end of inflation even when an adiabatic limit has not been reached. Before this limit is reached, $Z_{i}$ demonstrates significant variation. The extent to which $Z_{i}$ varies over the course of the last 55 e-folds of inflation changes between inflationary realisations according to both the realisation of the random coefficients and the path the inflationary trajectory takes in field space, but an example of the form of the variation is given in Fig.~\ref{fig:Z}, where the fields have been redefined such that the inflection point is at the origin. 

As the trajectory crosses the inflection point, $Z_{i}$ has a trough, then increases to zero as the trajectory leaves the plateau. As can be seen in Eq.~\eqref{Pzeta}, a peaking in $|Z_{i}|$ corresponds to a peak in $P_{\zeta\zeta}$, though the precise shape will also depend on the angular component which determines for how long the trajectory stays on the corresponding ridge in $Z_{i}$. The peak in $|Z_{i}|$ occurs due to $\epsilon^{c}$ in the denominator of Eq.~\eqref{eq:Z} reaching a minimum value and as such non-monotonic evolution is an inevitable consequence of the inflationary trajectory crossing an inflection point. The form of $Z_{i}$ also accounts for the final rise (the caterpillar's head) in the evolution of $P_{\zeta\zeta}$ since $Z_{i}$ turns up as a sum of quadratic terms in Eq.~\eqref{Pzeta}. This stage of the evolution is non-slow-roll but when comparing the evolution as given by $Z_{i}$ with our non-slow-roll transport code we found them to be in good agreement. 

\subsection{Why so blue?}

Recognising that all inflationary trajectories take place in the vicinity of an inflection point also tells us about the spectral index. Remembering that $Z_{i}\rightarrow0$ towards the end of inflation, Eq.~\eqref{eq:n} states that the final value of the spectral index (along with the other observables we discuss) will be determined purely in terms of horizon-crossing values. This is what is often referred to as `the horizon-crossing approximation' \cite{SA}.  Unless very close to the inflection point,  $\eta_{x}$ dominates Eq.~\eqref{eq:n}. When horizon-crossing takes place prior to the inflection point, $\eta_{x}>0$, the horizon-crossing approximation says that $n_{s}$ will necessarily be larger than 1. However if the situation arises where horizon-crossing takes place after the inflection point, $\eta_{x}<0$ and a red spectral index can be expected. This is what we find to be the case for approximately $16\%$ of our trajectories (for $N_{\rm F}=2$). The reason they are so rare is that 60 e-folds of inflation still need to take place and this is difficult to achieve before falling off the ledge. It turns out that for the red cases, the total number of e-folds was always at least ${\cal O}(100)$.  

Fig.~\ref{dtest} shows a sensitivity to the number of fields $N_{\rm F}$, where additional angular terms appear to act to redden the spectral index. This is not due to a larger proportion of trajectories with horizon-crossing occurring after the inflection point. If this was the case then there would also be an increase in the number of e-folds. Instead, the increased redness seems to be a direct result of the geometry of the conifold and can be accounted for by the separable approximation. Writing Eq.~\eqref{eq:n} in a more suggestive form
\begin{equation}
n_{\rm s}-1=-2\sum_{i=1}^{N_{\rm F}}\epsilon^{*}_{i}-\frac{4}{\sum_{i=1}^{N_{\rm F}}\frac{u_{i}^2}{\epsilon^{*}_{i}}}+2\frac{\frac{\eta^{*}_{x}}{\epsilon^{*}_{x}}u_{x}^2}{\sum_{i=1}^{N_{\rm F}}\frac{u_{i}^2}{\epsilon^{*}_{i}}},
\end{equation}
it is clear that additional angular terms act solely to redden the spectral index by suppressing the contribution from $\eta_{x}$ as well as increasing $\epsilon$.

\subsection{Trends in the running}

Fig.~\ref{scats2} shows a lower bound in the plot of the running against the spectral index. 
This bound turns out to be parabolic and approximately  proportional to $ -(n_{s}-1)^2$, suggesting that there could be a dominant term in $\eta_{x}^2$ among the contributions for the running. Interestingly, in the single-field case, terms in $\eta^2$ do not contribute to the running, $n^{\prime}=-24\epsilon^2 +16\epsilon \eta - 2 \xi$. However, as can be seen by following Eq.~\eqref{eq:runsr}, in the multifield case the situation changes. The terms from Eq.~\eqref{eq:runsr} in $\eta_{x}^2$ are:
\begin{equation}
n^{\prime} \supset -4\left(\frac{\sum_{i}\frac{\eta^{*}_{i}u_{i}^2}{\epsilon^{*}_{i}}}{\sum_{i}\frac{u_{i}^2}{\epsilon^{*}_{i}}}\right)^2+4\frac{\sum_{i}\frac{\eta^{*2}_{i}u_{i}^2}{\epsilon^{*}_{i}}}{\sum_{i}\frac{u_{i}^2}{\epsilon^{*}_{i}}}=
-4\frac{\frac{\eta^{*2}_{x}u_{x}^4}{\epsilon^{*2}_{x}}}{\left(\sum_{i}\frac{u_{i}^2}{\epsilon^{*}_{i}}\right)^2}+4\frac{\frac{\eta^{*2}_{x}u_{x}^2}{\epsilon^{*}_{x}}\sum_{j}\frac{u_{j}^2}{\epsilon^{*}_{j}}}{\left(\sum_{i}\frac{u_{i}^2}{\epsilon^{*}_{i}}\right)^2},
\end{equation}
which means that there is a residual $\eta_{x}^2$ cross term which dominates the expression. The parabolic lower bound is therefore intrinsically multifield.


\subsection{Evolution in $f_{\rm NL}$}

\begin{figure}[t]
\centering
\includegraphics[width=10cm]{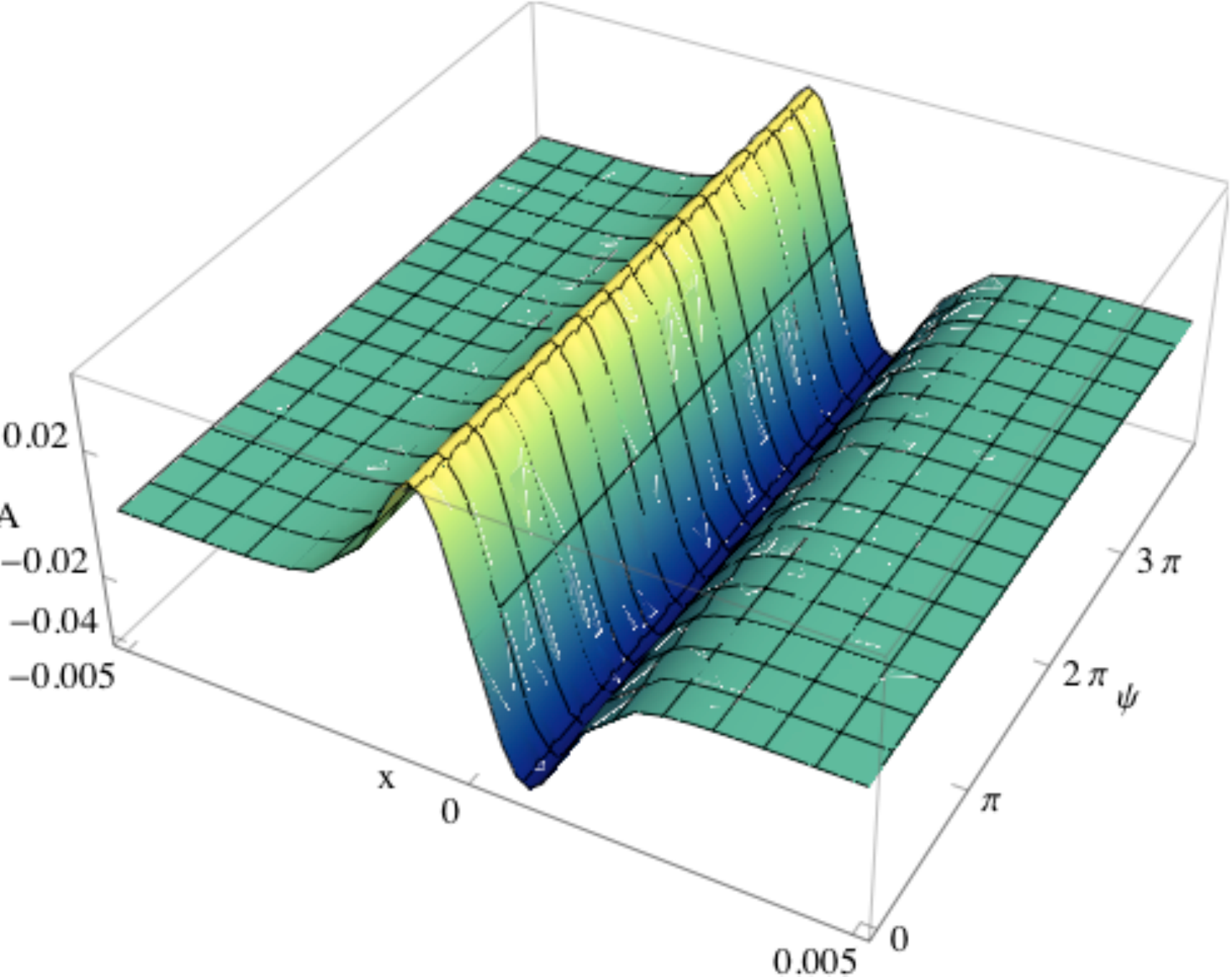}
\caption{ Example of the form of $A_{ji}$ for the separable model.}
\label{fig:A}
\end{figure}

The separable approximation fares less well in reproducing the behaviour seen for $f_{\rm NL}$. The radial acceleration parameter $\eta_{x}$ is the dominant contribution to Eq.~\eqref{eq:fnlsr}, so the horizon-crossing approximation gives $f_{\rm NL}<0$ for horizon-crossing prior to the inflection point and $f_{\rm NL}>0$ for post inflection-point horizon-crossing. Bringing this together with the previous discussion for the spectral index implies there should be a correlation between $n_{s}$ and $f_{\rm NL}$ such that a blue spectral index is accompanied by $f_{\rm NL}<0$ and vice versa. This is in fact the relation from single-field inflation, $f_{\rm NL}=-5(n_{s}-1)/12$~\cite{Maldy}. While Fig.~\ref{scats1} clearly shows this line, it is also clear that the majority of points deviate from this trend. In fact, the histogram for $f_{\rm NL}$ shows a tendency for $f_{\rm NL}>0$, while the discussion until now would imply the opposite.

Looking closely at particular evolutions of $f_{\rm NL}$, like the one in Fig.~\ref{fig:trajetc}, what we see is that during the non-slow-roll period at the end of inflation, the evolution of  $f_{\rm NL}$ experiences a rise, that typically forces the final value to be $>0$. This behaviour is not predicted by the separable approximation. This discrepancy could be the result of modelling the inflection point as being cubic when in fact the ledge in the D-brane potential tends to be much more severe, or more interestingly it could be an intrinsically non-slow-roll effect.

However, the slow-roll phase of the evolution does seem to be captured by the separable approximation. In addition to $Z_{i}$, there is contribution to the evolution $A_{ji}$ which, as shown in Fig.~\ref{fig:A}, has a peak and a trough. While the precise form varies between trajectories, this contribution gives rise to a peak and a trough in $f_{\rm NL}$, one above and one below the value $f_{\rm NL}$ takes towards the end of inflation.\\

 


\section{How predictive?}\label{AL}

A consideration of paramount importance in any model of inflation is at what point does $\zeta$ cease to evolve. In single-field models, $\zeta$ does not evolve on superhorizon scales but for any model with more than one field this tends not be the case. As we have already discussed, the criteria for whether or not evolution can persist is whether or not isocurvature modes are present. The limit in which these decay is referred to as the adiabatic limit. 
\begin{figure}[t]
\centering
\includegraphics[width=15cm]{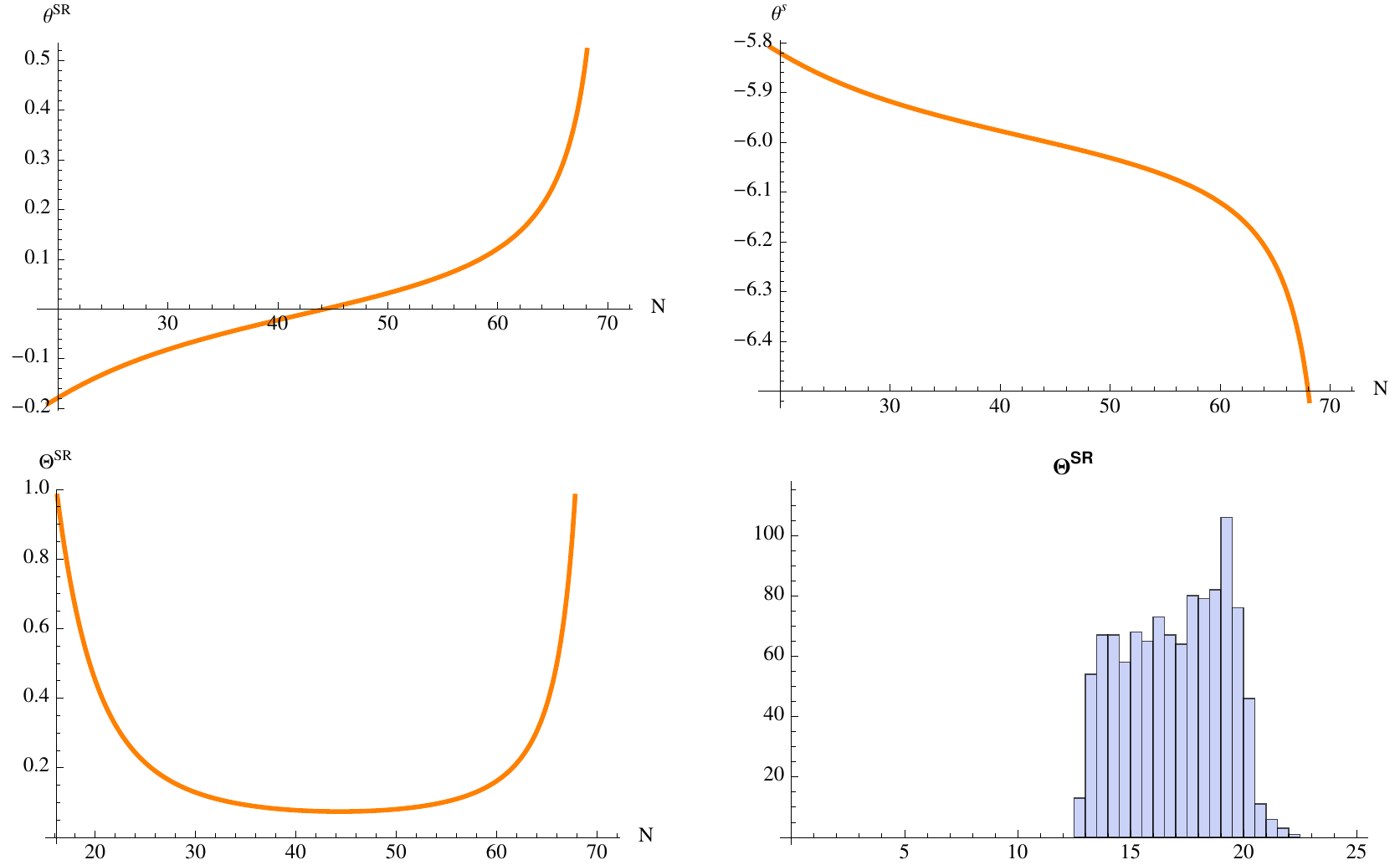}
\caption{Evolution of $\theta^{\rm SR}$ top left, $\theta^{\rm s}$ top right, and the dilation of the field bundle $\Theta^{\rm SR}$ bottom left for verse 20277. Bottom right is the histogram of the final field bundle widths for our two-field sample. The fields initially go through a region of focussing but all is undone as the trajectory falls off the ledge, resulting the persistence of isocurvature modes at the end of inflation. The histogram shows this to be true for all 1086 realisations, leading us to conclude the model is unpredictive without a description of reheating.}
\label{fig:theta}
\end{figure}

As already mentioned, one way to keep track of isocurvature modes is to monitor the dilation of the bundle of trajectories using Eq.~\eqref{eq:thetah}. In the adiabatic limit the bundle is a caustic, for which a necessary but not sufficient condition is $\theta\rightarrow -\infty$, or equivalently $\Theta\rightarrow 0$. Strictly speaking this limit cannot be reached during the slow-roll regime \cite{optics} but we can at least hope that isocurvature modes are exponentially suppressed.  More generally $\theta^{\rm SR}<0$ corresponds to a region of focussing and $\theta^{\rm SR}>0$ to dilation, while $\theta^{\rm s}>0$ and $\theta^{\rm s}<0$ represent divergence and convergence to the momenta slow-roll attractor respectively.

It is interesting to note that our results show exceedingly consistent behaviour. As shown by the histogram in Fig.~\ref{fig:theta}, in not one realisation did we find focussing to be occurring at the end of inflation. Rather, the process of falling off the ledge results in dilation of the bundle. This result is still present in the separable approximation and indeed we expect this result to apply to any model where inflation is terminated by falling off a ledge if sufficient focussing has not taken place previously. 

The downside of a lack of focussing at the end of inflation is that the model is technically unpredictive without knowledge of what takes place at the tip and the details of reheating. The plus side is that one should expect interesting evolution during subsequent periods. Since reheating is a non-linear process, this could give rise to interesting observational signatures. This is however beyond the scope of this paper.

\section{Summary}

In this paper we explored the multifield effects in D-brane inflation. To do this we made use of a particularly sophisticated model, originally developed in Ref.~\cite{Agarwal}, to include contributions from the bulk containing random coefficients. These contributions make the multifield nature of D-brane inflation explicit. Our aim was to use the transport equations \cite{MT1, MT2, us, optics} to study a large number of realisations of the potential, their resulting inflationary behaviours, and the consequent properties of $\zeta$. 
We had two main objectives with this endeavour, first to present distributions for the observable predictions of this specific model and second to analyse how the isocurvature modes behave towards the end of inflation to understand how reliable those predictions are. 

Despite the random contributions, the inflationary behaviour was very consistent across different realisations. Inflation was always found to have canonical kinetic behaviour and to take place in the vicinity of the inflection point, which constitutes a very small region of field space. We also always found a break of the slow-roll condition $|\eta|<1$ when the inflationary trajectory falls off the ledge after the inflection point. This particular characteristic led to the extension of transport methods to non-slow-roll cases \cite{us}. We also extended this method to compute the running of the spectral index. 



Regarding the predictions for observables, the amplitude of the power spectrum was found to be consistent with observation but unpredictive in so much as the spread in values was vast compared with the observational constraints. The spectral index tended to be blue but less so in the full six-field case. This fact, together with the negligible values found for the tensor-to-scalar ratio and the large spread in values of the spectral index running, moderately constrains the model given WMAP7 data. Without a reason to believe we are an atypical observer in our distribution, the soon-to-arrive data from Planck has the potential to put this model under considerable pressure. 

The non-Gaussianity parameter $f_{\rm NL}$ was found to be consistently small. 
In over 1000 inflationary realisations we found only 5 examples of potentially observable $f_{\rm NL}$. This further supports the recent discussions in Refs.~\cite{me2, Meyers, JDD} which suggests canonical multifield inflation is not expected to produce observable $f_{\rm NL}$.

Due to the remarkable consistency in the inflationary behaviour, it was possible to approximate the inflationary region of the potential as a sum-separable potential consisting of a polynomial inflection point in the radial direction and linear slopes in the angular directions. With this description all the observed characteristics of this model could be accounted for analytically except the final evolution in $f_{\rm NL}$. It was shown that the evolution of the amplitude of the power spectrum undergoes non-monotonic evolution whenever the trajectory traverses the inflection point and this will typically give rise to a blue tilt which, along with all other observable quantities of interest, can be calculated purely in terms of horizon-crossing values. It was also shown that additional angular terms act to make the tilt more red.

Regarding the reliability of these predictions, we made first steps towards a means of efficiently tracking isocurvature modes in non-slow-roll multifield models. Though we leave a complete description of this to a separate publication, we nevertheless found the bundle width to be exceedingly informative. Extending the ideas on the adiabatic limit discussed in Ref.~\cite{optics} to non-slow roll constitutes a considerable increase in complexity. We found it helpful to consider two distinct bundles, one for the fields and a second for the momenta. With this we were able to conclude that in not a single realisation was an adiabatic limit reached. We expect this result to hold for any model where inflation is terminated by falling off a ledge, provided sufficient focussing does not occur prior to this. This result renders the model technically unpredictive, since subsequent events such as reheating can modify the characteristics of $\zeta$.

\acknowledgments
We would particularly like to thank David Seery for numerous discussions and guidance throughout this work. In addition we would also like to thank David Mulryne for advice on the transport method and Liam McAllister and Gang Xu for help with the building of the potential. M.D and J.F would also like to thank Tic Toc cafe, Brighton, where quite possibly the majority of this work was done. J.F also gives thanks  to Leon Baruah for fixing his laptop at a critical stage of the writing up. M.D. was supported by FCT (Portugal), J.F. and A.R.L  were supported by the Science and Technology Facilities Council [grant numbers ST/1506029/1, ST/F002858/1, and ST/I000976/1] and A.R.L. by a Royal Society--Wolfson Research Merit Award. 


\pagebreak

\renewcommand{\topfraction}{0.99}

\part*{Erratum: Multifield consequences for D-brane inflation}

\addcontentsline{toc}{part}{Erratum: Multifield consequences for D-brane inflation}
\vskip2cm

\section{Curved field-space metric and massive modes}

In the analysis presented in our paper `Multifield consequences for D-brane inflation', we considered the field-space metric in Eq.~(2.4) to be the flat trivial metric, $g_{ij}=\delta_{ij}$. However, the inclusion of the curved conifold metric has several implications on the inflationary dynamics, and as such should be incorporated in our calculations. We thank Liam McAllister and Sebastien Renaux-Petel for pointing out this issue. 

In this case, the Lagrangian experienced by the D-brane is
\begin{equation}
{\cal L}=a^{3}\left(\frac{1}{2}T_{3}G_{ij}\dot{\phi}^{i}\dot{\phi}^{j} -V(\phi) \right)
\end{equation}
where $a$ is the scale factor and $T_{3}$ is the brane tension. For simplicity, the vector composed of the 6 brane coordinates in the throat --- radial and $\Psi$ --- is represented by $\phi$ and the vector $\Psi$ refers to the 5 angular dimensions, $\Psi={\theta_1,\theta_2,\phi_1,\phi_2,\psi}$.

The field-space metric $G_{ij}$ corresponds to the Klebanov--Witten geometry in which the non-compact conifold geometry is built over the five-dimensional $\left(SU(2) \times SU(2)\right)/U(1)$ coset space $T^{1,1}$. In this case \cite{CAO}, 
\begin{equation}
G_{ij}d\phi^id\phi^j=dr^{2}+r^{2}ds^{2}_{T^{1,1}}.
\end{equation}
where $r$ is the radial conical coordinate and 
\begin{equation}
ds^{2}_{T^{1,1}}=\frac{1}{9}\left(d\psi + \cos\theta_1d\phi_1+\cos\theta_2d\phi_2\right)^2+\frac{1}{6}\left(d\theta_1^2+\sin^2\theta_1d\phi_1^2\right)+\frac{1}{6}\left(d\theta_1^2+\sin^2\theta_1d\phi_1^2\right).
\end{equation}
With this new Lagrangian we repeated the experimental procedure described in \S3, and collected over 500 trajectories, with 6 active fields and $\Delta_{\rm MAX}=3$, which give rise to at least 63 $e$-folds of inflation.

It is usual that models with curved field-space metrics present massive modes around horizon crossing. To verify if this was the case in our model, we calculated the eigenvalues of the mass matrix. The mass matrix can be determined from the second-order perturbed action:
\begin{equation}
S_{(2)}=\int dt d^3x a^3\left(G_{ij}{\mathcal D}_tQ^i{\mathcal D}_tQ^j - \frac{1}{a^2}G_{ij}\partial_\mu Q^i \partial^\mu Q^j- M_{ij}Q^iQ^j \right),
\end{equation}
where ${\mathcal D}_t$ are covariant cosmic time derivatives, $Q^i$ are the covariant field perturbations  and 
\begin{equation}
M_{ij}=V;_{ij}-R_{iklj}\dot{\phi}^k\dot{\phi}^l-\frac{1}{a^3}{\mathcal D}_t\left[\frac{a^3}{H}\dot{\phi}_i\dot{\phi}_j\right]
\end{equation}
with semicolon referring to covariant derivatives. From this action we can recover the Klein--Gordon equation for the field perturbations $Q^i$
\begin{equation}
{\mathcal D}_t^2Q^i+3H{\mathcal D}_tQ^i+\frac{k^2}{a^2}Q^i+M^i_jQ^j=0.
\end{equation}

\begin{figure}[t]
\centering
\includegraphics[width=13.5cm]{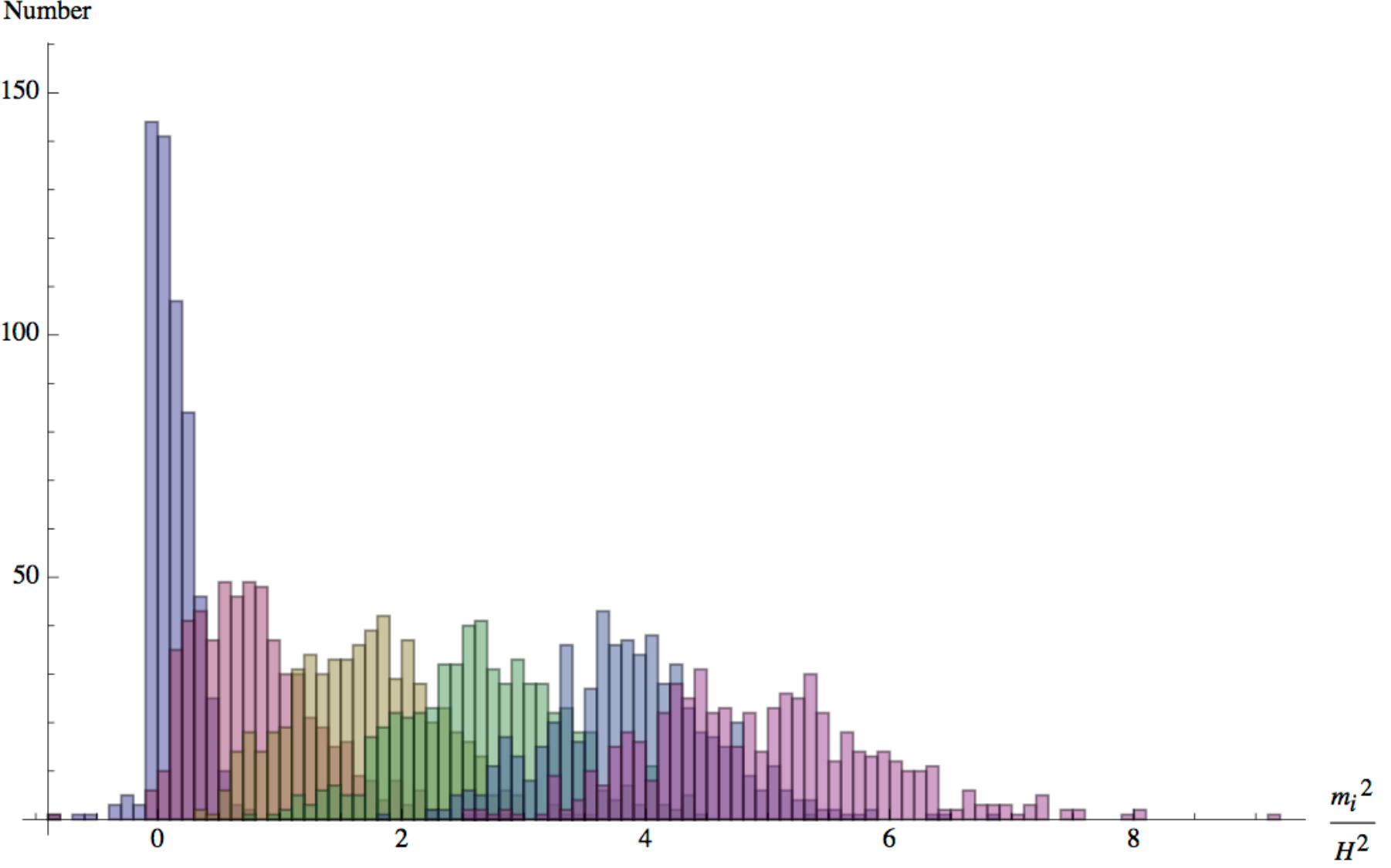}
\caption{The eigenvalues of the mass matrix at horizon crossing for all the successful realisations. We can see that there is consistently one light quantum mode while all the others are heavy.}
\label{lotsofmasses}
\end{figure}

For our sample of trajectories we consistently found one of the directions to be light while the other quantum modes have masses of order of the Hubble scale at horizon exit, in agreement with Ref.~\cite{liam2}. The results are shown in Fig.~\ref{lotsofmasses}. The presence of heavy modes required the development of a new computational tool for the calculation of $\zeta$. The problem relates to the initial condition at horizon exit necessary for implementing the separate universe assumption. In our previous analysis, the standard light-field Gaussian approximation at horizon exit was sufficient --- $\Sigma^{\phi\phi}=G_{ij}H^2/(2\pi)^2$. But now this approximation does not hold. An analytic solution of the full Klein--Gordon equation is impossible to achieve and for this reason we developed with our collaborators David Mulryne and David Seery an efficient method to deal with this issue numerically \cite{us,mulry}. Here, we only present the analysis for the two-point function of $\zeta$, leaving a detailed study of the three-point function for a future publication.

\section{New techniques}

In this section, our aim is to sketch the extensions to the transport method (\S4 of our paper) used in the new analysis of the D-brane model. First, we will briefly present the argument of Elliston \textit{et al.}\ \cite{newjoe} applied to a non-slow-roll setup, and then we will introduce the idea of quantum transport on subhorizon scales. We will not develop these descriptions in great detail here; we refer the reader to Refs.~\cite{mulry,us} for a complete description.  

If inflation is driven by a system of several scalar fields whose field-space metric $G_{ij}$ is different from $\delta_{ij}$, the field perturbations $\delta\phi$ of \S4 are no longer covariant objects. To use the transport method in this case, we need to reformulate it in terms of the above covariant perturbations $Q^i$.  In the same spirit as \S4 we can define a vector ${\mathcal Q}^\alpha \equiv \{Q^i, {\mathcal D}_N Q^i\} \equiv \{Q^i, P^i\}$. 
The foundation of the transport formulation is the deviation equation which determines the evolution of perturbations according to background expansion tensors. For the new covariant perturbations this looks like
\begin{equation}
{\mathcal D}_N{\mathcal Q}^\alpha= w^\alpha_{\;\;\beta}{\mathcal Q}^\beta+\frac{1}{2}w^\alpha_{\;\;\beta\gamma}{\mathcal Q}^\beta{\mathcal Q}^\gamma + \cdots
\end{equation}
For the computation of the two-point function, the first-order term is sufficient. The expansion tensor $w^{\alpha}_{\;\;\beta}$ can be read off directly from the Klein--Gordon equation for $Q^i$ given above by taking the superhorizon limit $k^2/a^2\ll1$. If we change the time coordinate to be $N$ rather than cosmic time, and assuming that we are well outside the horizon,
\begin{align}
\mathcal{D}_N Q^i=&P^i\\
\mathcal{D}_N P^i=&\left[-\frac{V;^i_{\	j}}{H^2}+\frac{V,^i V,_{j}}{H^3V}+R^i_{\	lmj}p^lp^m \right]Q^j\\ \nonumber
+& \left[\frac{V,^i p_{j}}{(3-\epsilon)H^2}+p^ip_j+(\epsilon-3)\delta^i_j \right]P^j. \nonumber
\end{align}
The expansion tensor $w^\alpha_{\;\;\beta}$ contains all the information on the superhorizon evolution of the power spectra $\Sigma^{\alpha\beta}$ associated to $\langle {\mathcal Q}^\alpha{\mathcal Q}^\beta \rangle$ 
\begin{align}
\label{nontrivial}
&{\mathcal D}_N \Sigma^{\alpha\beta}=w^{\alpha}_{\	\	\mu}\Sigma^{\mu\beta}+w^{\beta}_{\	\	\mu}\Sigma^{\mu\alpha}+\cdots
\end{align}
The power spectra $\Sigma^{\alpha\beta}$  relate to the power spectrum of $\zeta$ by the usual gauge transformations described in \S4.\\

This method describes the transport of field perturbations on superhorizon scales using the separate universe assumption. The question to ask now is what are the initial conditions for this method at horizon exit. Since it is impossible to deal with this question analytically, correlation functions need to be transported from subhorizon scales up to the end of inflation, so that the full evaluation is numerical. Usually, a numerical implementation of perturbation theory is computationally very intensive. This is related to the oscillatory nature of the perturbations as wave functions. However, by only keeping track of the evolution of correlators, the calculation becomes much lighter. 

In order to do this, we extended the transport technique to a quantum era on subhorizon scales. In this case, if the evaluation is chosen to start early enough, such that the perturbations are much smaller than the horizon, one can assume that the initial conditions are established in flat Minkowski space-time, where correlation functions are well known. 

To quantise transport, the field perturbations and their momenta should be interpreted as operators in Fourier space related by the usual commutation relations. In $k$ space it is helpful to use a DeWitt notation where indices are primed to indicate the Fourier space scale dependence:
\begin{equation}
\widehat{\mathcal Q}^{\alpha'}=\widehat{\mathcal Q}^\alpha(k_\alpha)
\end{equation}
where $k_\alpha$ is the scale associated with the perturbation of index $\alpha$ (with no sum on $\alpha$).
This operator has an evolution equation like
\begin{equation}
\label{nontrivialwquantum}
{\mathcal D}_N\widehat{\mathcal Q}^{\alpha'}= w^{\alpha'}_{\	\	\beta'}\widehat{\mathcal Q}^{\beta'}+\frac{1}{2}w^{\alpha'}_{\	\	\beta'\gamma'}\widehat{\mathcal Q}^{\beta'}\widehat{\mathcal Q}^{\gamma'} + \cdots
\end{equation}
where
\begin{equation}
w^{\alpha'}_{\;\;\;\beta'}=(2\pi)^3w^{\alpha}_{\;\; \beta}\,\delta(k_\alpha-k_\beta) .
\end{equation}
For the quantum subhorizon evolution the expansion tensor is not determined by the background only. It can be recovered by looking at the second-order perturbed action computed above. In this case, the quantum expansion tensor $w^\alpha_{\;\;\beta}$ only gets modified by the extra term containing information on the (comoving) scale $k$ of the perturbations:
\begin{equation}
w^\alpha_{\;\;\beta} \supset -\frac{k^2}{a^2H^2}\delta^\alpha_{\beta}.
\end{equation}
The evolution equation for $\Sigma^{\alpha\beta}$ follows from the Ehrenfest theorem 
\begin{equation}
\frac{d\langle\widehat{O}\rangle}{dt}=\Big\langle-i\left[\widehat{O},\widehat{H}\right]\Big\rangle,
\end{equation}
which is the direct equivalent to the evolution of classical expectation values. For simplicity, one can choose to work with the symmetrised $\Sigma^{\alpha\beta}$, which corresponds to the real part of the two-point correlator; as perturbations become classical, the imaginary part decays which means that only the evolution of the real part is necessary to compute observables. The transport equation for this symmetrised $\Sigma^{\alpha\beta}$ is equivalent to the one on superhorizon scales:
\begin{equation}
{\mathcal D}_N \Sigma^{\alpha\beta}=w^{\alpha}_{\;\;\mu}\Sigma^{\mu\beta}+w^{\beta}_{\;\;\mu}\Sigma^{\mu\alpha}+\cdots
\end{equation}

With these equations we were able to compute the perturbations for our sample of trajectories. We started our computation 8 $e$-folds of inflation before horizon exit, which is early enough to find the perturbations well inside the horizon where Minkowski initial conditions hold. We computed the observables for the perturbation associated with the scale which crossed the horizon 55 $e$-folds before the end of inflation.

\section{Results}

\subsection{Observables}

We now present the results from our new analysis. The outcome is distributions for the values of the cosmological parameters associated to the two-point function of perturbations: amplitude $P_{\zeta\zeta}$, spectral index $n_{s}$, and tensor-to-scalar ratio $r$. We compared these with constraints from observations; all constraint contours are $95\%$ confidence limits using the WMAP 7-year data release combined with baryonic acoustic oscillations and supernov\ae \  data \cite{17}.  

We concluded that the inclusion of the curved field-space metric, even though it induces large masses in 5 of the 6 quantum modes active during inflation, does not qualitatively change the results for observables that we obtained with a trivial field-space metric.

%

As can be seen in Fig.~\ref{hists}, the histogram of $P_{\zeta\zeta}$ has a smooth maximum at around $10^{-9}$, in agreement with observations (the WMAP value is $\sim 2.5 \times 10^{-9}$ \cite{17}) and with our original computation. This is not surprising as the overall magnitude of the potential is determined by the scale $\mu^{4}$, which in turn is set by our choice of the throat length $r_{\rm UV}$.

The spectral index still presents a peak around $n_{s}=1$ but it is less dramatic than in our previous computation, as seen in Fig.~\ref{hists}. As before, two different populations can be identified, $73\%$ of the realisations with $n_{s} \geq1$ and $27\%$ with $n_{s}<1$.

The tensor-to-scalar ratio is always extremely small, as it is related to the slow-roll parameter $\epsilon$ that remains $\ll 1$ throughout inflation. This can be clearly seen in Fig.~\ref{scats2}.

\begin{figure}[t]
\centering
\includegraphics[width=16cm]{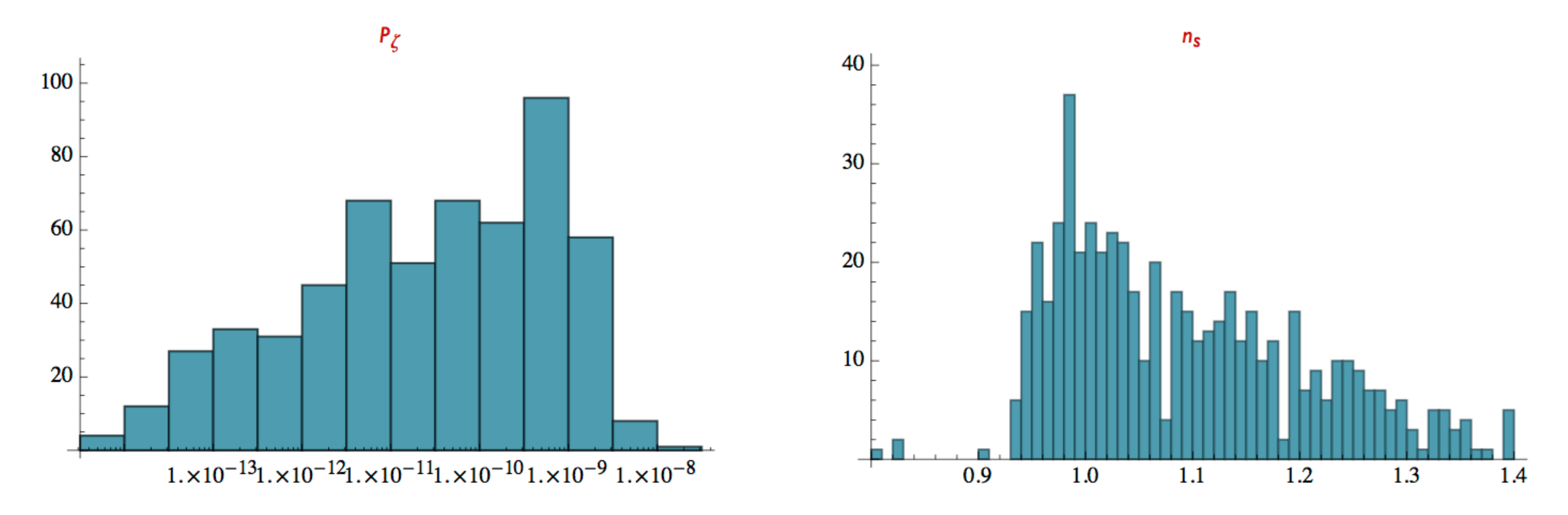}
\caption{Distributions for the amplitude of the power spectrum $P_{\zeta\zeta}$, left and scalar spectral index $n_{s}$, right.}
\label{hists}
\end{figure}

Imposing observational constraints on the distributions, as can be seen in Fig.~\ref{scats2}, excludes the majority of the trajectories. Almost all the realisations with red tilt are in accordance with observational constraints, such that $\sim 20\%$ of the total sample is in agreement with data.
A further constraint is imposed by requiring the correct amplitude of the scalar power spectrum, $(2.5 \pm 0.1) \times 10^{-9}$ \cite{17}. Combining all constraints we obtained only three realisations in total concordance with observations in the full sample of 564 cases. 
As discussed in the original article, this is not a worrying result as the distribution of $P_{\zeta\zeta}$ does not show a sharp peak.\\

\begin{figure}[t]
\centering
\includegraphics[width=15cm]{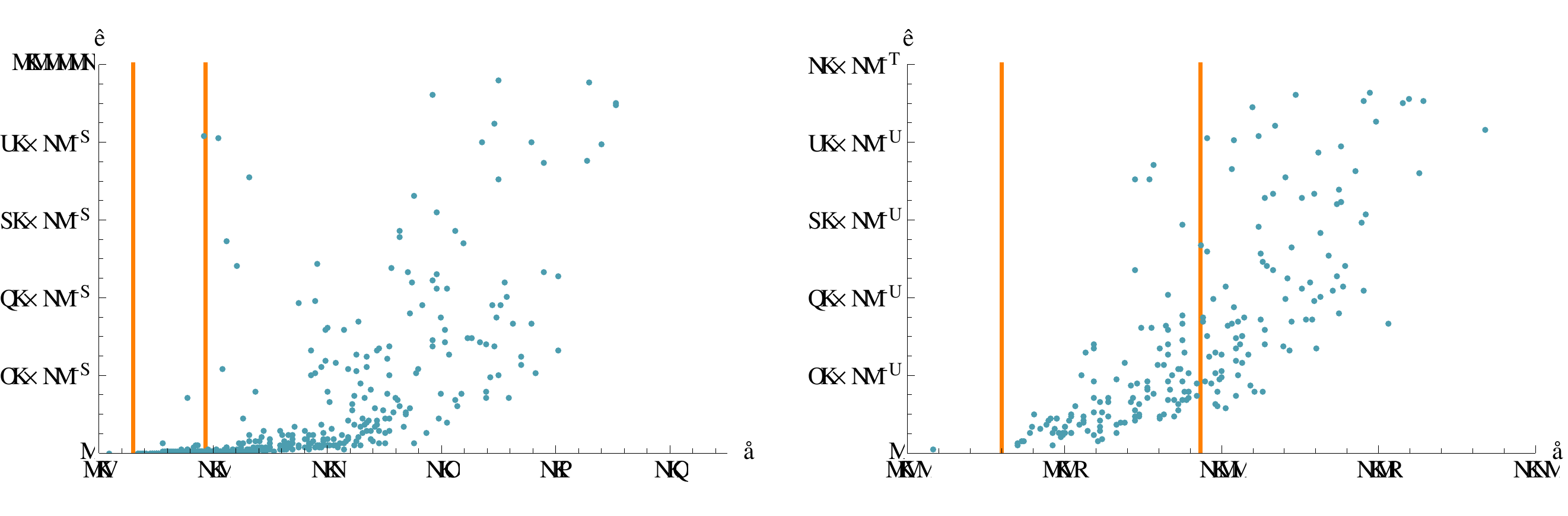}
\caption{Plot of the values of $n_{s}$ versus  $r$. The right panel greatly expands the vertical scale. The orange lines indicate the $95\%$ confidence limits using WMAP data. }
\label{scats2}
\end{figure}

\begin{figure}[t]
\centering
\includegraphics[width=14.3cm]{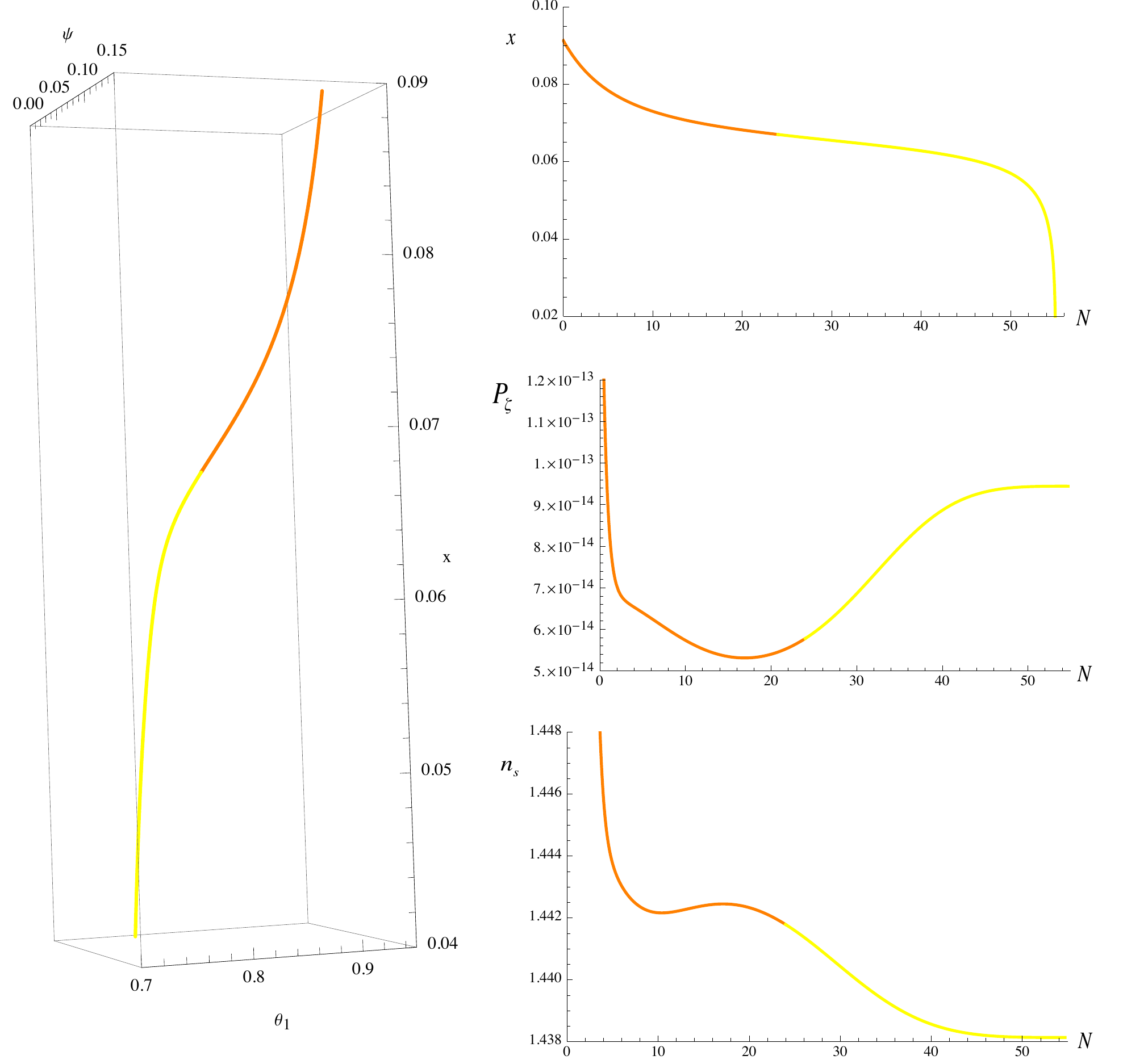}
\caption{The dynamical behaviour of Verse 79856. The left plot shows the projection of the trajectory across $x$, $\theta_1$ and $\psi$. The top-right panel shows the trajectory in $x$, whereas the middle and bottom panel show the superhorizon evolutions of $P_{\zeta\zeta}$ and $n_s$. Orange and yellow approximately indicates before and after the inflection point, respectively.}
\label{fig:trajetc}
\end{figure}

As in our original computation, all trajectories are essentially of the same type --- inflection-point inflation. Most of the inflation occurs in a small sub-region of the conifold (typically $0.02< x< 0.09$) in the vicinity of an inflection point in the radial direction. Looking closely at one particular representative trajectory, Verse  79856, this is evident. 

In the top right panel of Fig.~\ref{fig:trajetc}, the trajectory in the radial coordinate is plotted in two different colours to highlight the approximate position of the inflection-point -- orange before and yellow after it. The left panel shows the inflationary trajectory projected over three of the six directions -- radial,  $\theta_1$ and $\psi$. The trajectory evolves from top to bottom and it is easy to see that around the inflection point it undertakes turns in the angular directions. The turns distinguish this trajectory from single-field inflection-point inflation and, as discussed, have consequences in the statistics of $\zeta$. In fact, it is possible to see in Fig.~\ref{fig:trajetc} --- right middle and bottom panels --- that the values of $P_{\zeta\zeta}$ and $n_s$ undergo superhorizon evolution around these turns in field space.

\begin{figure}[t]
\centering
\includegraphics[width=9cm]{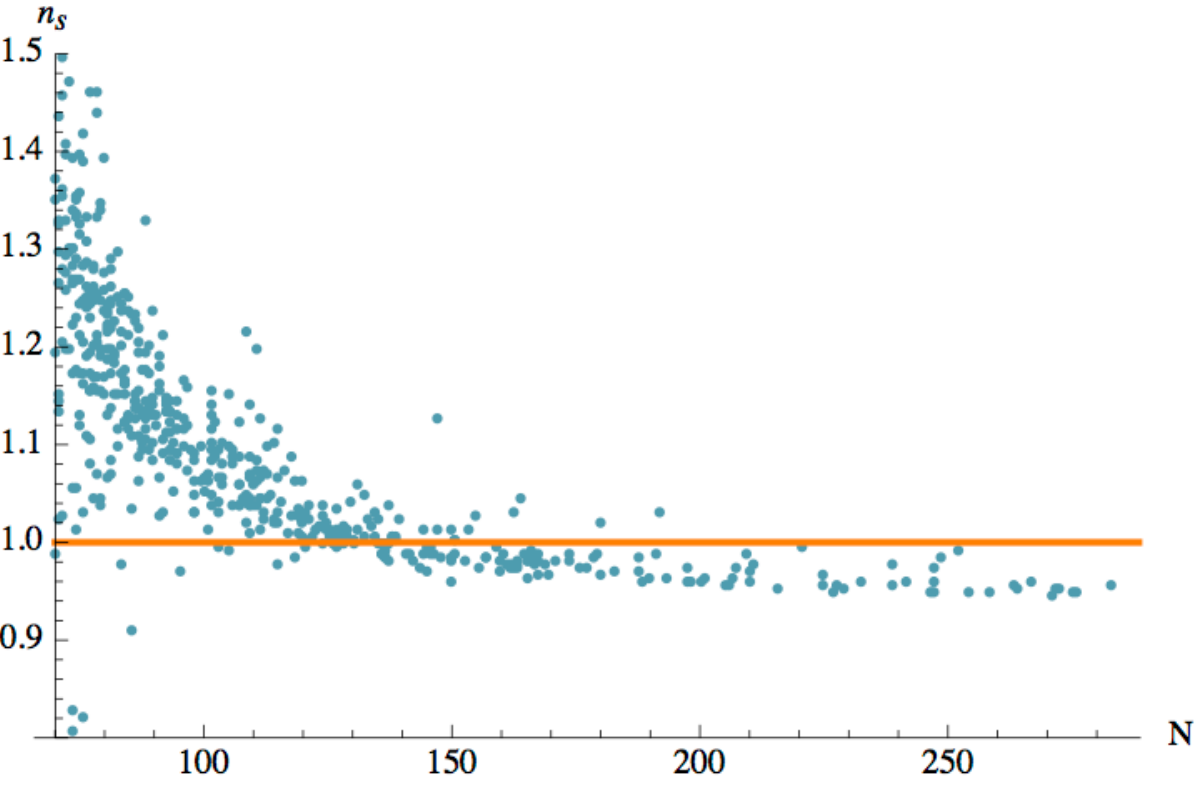}
\caption{The value of the spectral index plotted against the total number of $e$-folds. The orange line represents the $n_s=1$ cut. It is possible to see how $n_s<1$ implies $N \gtrsim 120$.}
\label{nsnend}
\end{figure}

As in our previous analysis, the populations with red and blue spectral index could be understood by the position of horizon crossing relative to the inflection point. In fact, it is straightforward to see that whenever horizon exit occurs before the inflection point, the spectral index is bigger than one. For the spectral index to be smaller than one, a dominant negative $\eta$ contribution is required, which implies horizon exit after the inflection point. The latter is harder to achieve, which explains the small proportion of inflationary trajectories with red tilt; a trajectory that gives rise to 55 $e$-folds in the yellow region needs to have a much larger total number of $e$-folds. Roughly, one would expect it to give rise to at least twice 55. Indeed, this rough estimation is confirmed by Fig.~\ref{nsnend}, where the value of the spectral index is plotted against the total number of $e$-folds.

\subsection{How predictive?}\label{AL}

The biggest discrepancy between our two analyses concerns the approach to an adiabatic limit. As stated in \S7, one way to keep track of isocurvature modes is to monitor the dilation of the bundle of trajectories.

When the number of fields is larger than one, the bundle of trajectories is a tensor of rank $n>1$. In this case, the condition $\theta^{\rm SR} \rightarrow -\infty$ is not sufficient to ensure the bundle is reaching a caustic, becoming an object of rank 0. In fact, $\theta^{\rm SR}$ would experience this behaviour whenever the bundle becomes a tensor of rank $n-1$. In other words, a `spherical' bundle could focus to a `sheet' rather than a point; if this happens, isocurvature modes are not suppressed \cite{optics}. Therefore it is hard to make absolute statements.

In the D-brane model, as can be seen in Fig.~\ref{fig:theta}, we consistently found across all the trajectories and throughout the full inflationary period $\theta^{\rm SR}<0$, which means the bundle of trajectories is focusing. To ensure this corresponds to reaching an adiabatic limit we have to make sure the bundles are reaching caustics. In this model, where we are near what can be considered quasi-single-field inflation, with one direction light and all the others heavy, there is no dynamical reason for the bundle to be focusing in any other way than by exponentially suppressing the isocurvature modes.
Therefore we can consider that this model does not raise problems of predictiveness related to persistence of isocurvature through reheating, in agreement with Ref.~\cite{liam2}.

\begin{figure}[t]
\centering
\includegraphics[width=15cm]{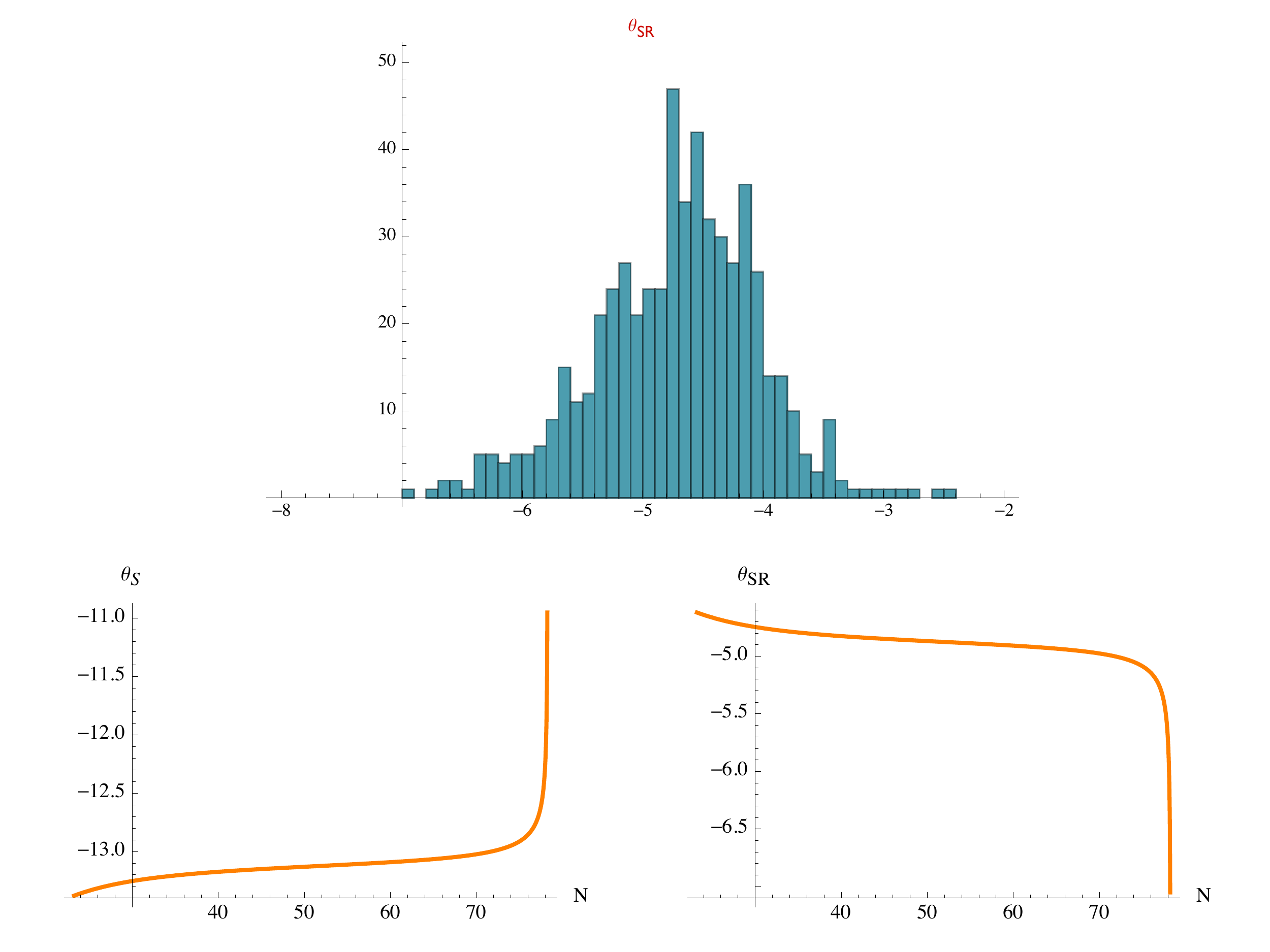}
\caption{On top, histogram of the final field bundle widths for our sample. On the bottom, evolution of $\theta^{\rm s}$ left and $\theta^{\rm SR}$ right for verse 79856.} 
\label{fig:theta}
\end{figure}

\section{Conclusions}

In this erratum we repeat our calculations for the predictions of the D-brane model presented in  `Multifield consequences for D-brane inflation' taking into account the curved field-space metric in the conifold. We performed this analysis for the power spectrum of curvature perturbations only, leaving the study of bispectrum for a future publication. The inclusion of curvature induces large masses in 5 of our 6 active quantum modes, in agreement with Ref.~\cite{liam2}, which has consequences for the dynamics of inflation. To perform this computation we used extensions of the transport method which allow for curved field space metrics and quantum subhorizon evolution. 

The results of our new analysis are in agreement with Ref.~\cite{liam2} and qualitatively identical to our original discussion, except regarding the suppression of isocurvature modes before the end of inflation, or, in other words, the reaching of an adiabatic limit. Our original conclusion, that the adiabatic limit was commonly not achieved, is not supported by our corrected analysis, where we find that the adiabatic limit is approached in all cases studied.

\section*{Acknowledgements}

We thank Daniel Baumann, Stephan Huber, Liam McAllister, David Mulryne, Sebastien Renaux-Petel, and David Seery for discussions relating to this erratum.

\end{document}